\documentclass[final,5p,times,twocolumn]{elsarticle}

\usepackage{amssymb,graphicx,amsmath, amsthm,verbatim}
\numberwithin{equation}{section}
\numberwithin{figure}{section}
\numberwithin{table}{section}

\newcommand\negspcfive{\!\!\!\!\!}%

\journal{Astroparticle Physics}

\begin{document}
\begin{frontmatter}
\title{Dark Matter and Fundamental Physics with the Cherenkov Telescope Array}

\author[uab]{M.~Doro\fnref{corr}}
\author[stockholm,kva]{J.~Conrad\fnref{corr}}
\author[southampton]{D.~Emmanoulopoulos}
\author[masc]{M.~A. S\`anchez-Conde}
\author[madrid]{J.A.~Barrio} 
\author[berlin]{E.~Birsin}
\author[lpnhe]{J.~Bolmont}
\author[cea]{P. Brun}
\author[sergio]{S.~Colafrancesco}
\author[connell]{S.~H.~Connell}
\author[madrid]{J.L.~Contreras}
\author[durham]{M.K.~Daniel}
\author[mattia]{M.~Fornasa}
\author[uab]{M.~Gaug}
\author[cea]{J.F. Glicenstein}
\author[ifae,multidark]{A.~Gonz\'alez-Mu\~noz}
\author[madrid]{T.~Hassan} 
\author[hamburg]{D.~Horns}
\author[lpnhe]{A.~Jacholkowska}
\author[erlangen]{C.~Jahn}
\author[mazini]{R.~Mazini}
\author[madrid]{N.~Mirabal} 
\author[ifae]{A.~Moralejo}
\author[cea]{E.~Moulin}
\author[madrid]{D.~Nieto} 
\author[stockholm]{J.~Ripken}
\author[UiB]{H.~Sandaker}
\author[berlin]{U.~Schwanke}
\author[berlin]{G.~Spengler}
\author[stamerra]{A.~Stamerra}
\author[cea]{A.Viana}
\author[hamburg]{H.-S.~Zechlin}
\author[stockholm]{S.~Zimmer}
\author[]{for the CTA collaboration.}

\fntext[corr]{Sent off-print requests to Michele Doro
  (\texttt{michele.doro@uab.cat}) and Jan Conrad (\texttt{conrad@fysik.su.se})}

\address[madrid]{Universidad Complutense de Madrid, E-28040 Madrid,
  Spain} 
\address[berlin]{Institut f\"ur Physik, Humboldt-Universit\"at zu
  Berlin, Newtonstr. 15, D 12489 Berlin, Germany} 
\address[lpnhe]{LPNHE, Universit\'e Pierre et Marie Curie Paris 6,
  Universit\'e Denis Diderot Paris 7, CNRS/IN2P3, 4 Place Jussieu,
  F-75252, Paris Cedex 5, France}
\address[cea]{CEA, Irfu, Centre de Saclay, F-91191 Gif-sur-Yvette, France}
\address[sergio]{School of Physics, University of the Witwatersrand,
  Johannesburg Wits 2050, South Africa \\
\& INAF - Osservatorio
  Astronomico di Roma, via Frascati 33, I-00040 Monteporzio, Italy.}
\address[connell]{University of Johannesburg, Johannesburg, South Africa}
\address[stockholm]{Oskar Klein Centre for Cosmoparticle Physics, 
Department of Physics,
  Stockholm University, Albanova, SE-10691 Stockholm, Sweden} 
\address[kva]{K. \& A. Wallenberg Research fellow of the Royal Swedish
  Academy of Sciences} 
\address[durham]{University of Durham, Department of Physics, South
  Road, Durham DH1 3LE, U.K.} 
\address[uab]{Universitat Aut\`onoma de Barcelona, Bellaterra, E-08193
  Barcelona, Spain}
\address[southampton]{Physics and Astronomy, University of
  Southampton, SO17 1BJ Southampton, United Kingdom}
\address[mattia]{Instituto de Astrofısica de Andalucıa (CSIC),
  E-18080 Granada, Spain \& Multidark Fellow} 
\address[ifae]{Institut de F\`isica d'Altes Energies (IFAE),
  Universitat Aut\`onoma de Barcelona, E-08193 Bellaterra (Barcelona),
  Spain} 
\address[hamburg]{Universit\"at Hamburg, Institut f\"ur
  Experimentalphysik, Luruper Chaussee 149, D 22761 Hamburg, Germany} 
\address[erlangen]{Universit\"at Erlangen-N\"urnberg, Physikalisches
  Institut, Erwin-Rommel-Str. 1, D 91058 Erlangen, Germany} 
\address[mazini]{Institute of Physics, Academia Sinica, Taipei 11529, Taiwan}
\address[masc]{Instituto de Astrof\`isica de Canarias, E-38205 La
  Laguna, Tenerife, Spain \\
\& Departamento de Astrof\`isica, Universidad de La
  Laguna (ULL), E-38205 La Laguna, Tenerife, Spain \\
  \& SLAC National Laboratory and Kavli Institute for
  Particle Astrophysics and Cosmology, 2575 Sand Hill Road, Menlo
  Park, CA 94025, USA} 
\address[UiB]{University of Bergen, Bergen, Norway}
\address[stamerra]{Department of Physics, University and INFN Siena, I-53100 Siena, Italy}

\begin{abstract}
The Cherenkov Telescope Array (CTA) is a project for a next-generation
observatory for very high energy (GeV--TeV) ground-based gamma-ray
astronomy, currently in its design phase, and foreseen to be operative
a few years from now. Several tens of telescopes of 2--3 different
sizes, distributed over a large area, will allow for a sensitivity
about a factor 10 better than current instruments such as H.E.S.S, MAGIC and
VERITAS, an energy coverage from a few tens of GeV to several tens of
TeV, and a field of view of up to 10~deg.  In the following study, we
investigate the prospects for CTA to study several science questions
that can profoundly influence our current knowledge of fundamental
physics. Based on conservative assumptions for the performance of the
different CTA telescope configurations currently under discussion, we
employ a Monte Carlo based approach to evaluate the prospects for
detection and characterisation of new physics with the array. 

\noindent
First, we discuss CTA prospects for cold dark matter searches,
following different observational strategies: in dwarf satellite
galaxies of the Milky Way, which are virtually void of astrophysical
background and have a relatively well known dark matter density; in
the region close to the Galactic Centre, where the dark matter density
is expected to be large while the astrophysical background due to the
Galactic Centre can be excluded; and in clusters of galaxies, where
the intrinsic flux may be boosted significantly by the large number of
halo substructures. The possible search for spatial signatures,
facilitated by the larger field of view of CTA, is also discussed.
Next we consider searches for axion-like particles which, besides being
possible candidates for dark matter may also explain the unexpectedly
low absorption by extragalactic background light  of gamma-rays from very
distant blazars. We establish the axion mass range CTA could probe through
observation of long-lasting flares in distant sources. Simulated
light-curves of flaring sources are also used to determine the
sensitivity to violations of Lorentz Invariance by detection of the
possible delay between the arrival times of photons at different
energies. Finally, we mention searches for other exotic
physics with CTA.
\end{abstract}


\end{frontmatter}

%

\clearpage
\section*{Introduction}
The Cherenkov Telescope Array (CTA)~\cite{CTAConsortium:2010a} will be
an advanced facility for ground-based gamma-ray astronomy in the
GeV--TeV regime. 
Compared to the
current generation of Imaging Atmospheric Cherenkov Telescopes (IACT)
e.g. H.E.S.S., MAGIC and VERITAS\footnote{Respectively
  www.mpi-hd.mpg.de/hfm/HESS/,\newline
  wwwmagic. mppmu.mpg.de/ and 
  veritas.sao.arizona.edu/}, CTA will feature substantial
improvements. It will cover over 3 decades in energy, from a few tens
of GeV up to several tens of TeV. At both ends of this range,
interesting new physics is expected, and in addition, a larger energy
coverage will provide a bigger lever arm for spectral studies.  Above
1 TeV, the
field of view (FOV) will be up to 10~deg i.e. over a factor of 2 larger than
that of current instruments. CTA is currently planned to have a Southern
hemisphere site and  Northern hemisphere site. This fact together
with the large
FOV of the telescopes in both installations will likely
enable CTA to provide the first extended gamma-ray maps of the sky in the TeV region. The improved energy and
angular resolution will enable more precise spectral and morphological
observation. This will be achieved by deploying several tens of
telescopes of 2--3 different sizes over an area of several square
km. CTA will be operated as an open observatory, with improved data
dissemination among the world-wide scientific community and a
substantial fraction of the total observation time devoted to guest proposals.

The search for new physics beyond the Standard Model (SM) of particle
physics is among the key science drivers of CTA along with the
understanding of the origin of high-energy gamma-rays and of the
physics of cosmic ray acceleration in galactic and extragalactic
objects. Several such fundamental
physics issues are examined here --- the nature of cold dark matter,
the possible existence of axion-like particles, and expected
violation of Lorentz Invariance by quantum gravity effects. Search
strategies for cosmic tau neutrinos, magnetic monopoles and follow-up
observations of gravitational waves, are also discussed.

The CTA array performance files and analysis algorithms are
extensively described in \citet{Bernloher:2012}. Eleven array
configurations ($A$\ldots$K$) were tested for the Southern hemisphere and two
($NA$, $NB$) for the Northern hemisphere \citep[Table 2]{Bernloher:2012}. The
simulations were made at an altitude of 2000 m and at 70 deg elevation.
Arrays $E$ and $I$ are considered balanced layouts in terms of performance
across the energy range. Arrays $A$, $B$, $F$ and $G$ are more focused to
low-energies, and arrays $C$, $D$ and $H$ to high energies. $NB$ is a higher
energy alternative to $NA$. Their point-source sensitivity is compared
in \citep[Fig.~7]{Bernloher:2012}. The arrays comprise different number of
telescopes of three different sizes: the Large Size Telescope (LST,
23~m diameter), the Medium Size Telescope (MST, 12~m diameter) and Small
Size Telescope (SST, 6~m diameter) \citep[Table 1]{Bernloher:2012}.
One of the goals of this
study was to compare different array configurations for the
specific scientific case. While in some
cases all CTA configurations are compared against each other, in
others only benchmarks array $B,\,C$ and $E$ are considered, as 
representative arrays that maximize the performance at low-energy,
high-energy and in the full-range, respectively. Except for galaxy cluster studies and Galactic halo
studies, where extended or diffuse MC simulations are used, in all
other cases point-like MC simulations are used. This is the first
time that realistic estimates of the prospects of detection for CTA are
presented for such searches. An optimised event
selection procedure and a dedicated analysis ought to improve
on our conservative expectations. 
Previous studies often relied on too optimistic sensitivities,
especially at low energies ($<100$ GeV); publicly available effective
areas for a subset of configurations
\citep{CTAConsortium:2010a,Bernloher:2012} are now accurate and can be
used to infer CTA sensitivities for point-like sources.\\

This contribution is structured as follows: 
\begin{list}{--}{\itemsep=0pt}
\item In Section~\ref{sec:dm}, we explore different possible scenarios
  for detection of cold dark matter particle signatures in
  observations of: dwarf satellite galaxies of the Milky Way
  (Section~\ref{sec:dwarf}), clusters of galaxies
  (Section~\ref{sec:gc}) and the Galactic halo
  (Section~\ref{sec:halo}). We also study anisotropies in the diffuse
  gamma-ray background as a signature of dark matter
  (Section~\ref{sec:spatial}).
\item In Section~\ref{sec:axions}, we discuss the scientific case for
  axion-like particles, and make predictions for detection from
  observation of blazars at different distances and with different
  flare durations.
\item In Section~\ref{sec:liv}, we compare the capacity of all planned CTA
  arrays to constrain high energy violations of Lorentz Invariance, relative to
  current limits. 
\item In Section~\ref{sec:exotic} we discuss qualitatively three more
  cases: the observation of air showers from $\tau-$leptons
  emerging from the Earth's crust 
  (Section~\ref{sec:tau}), the capability to identify magnetic
  monopoles as bright emitters of Cherenkov light in the atmosphere
  (Section~\ref{sec:monopoles}) and some consideration about
  multi-wavelength gravitational wave campaigns
  (Section~\ref{sec:gw}).
\end{list}

Given the wide variety of physics issues considered in this
contribution, an introduction to the individual physics case is
presented in each section for easier readability. The reader can
find an overall summary and closing remarks in
Section~\ref{sec:conclusion}.

%
%
%

\section{Cold Dark Matter Particle searches\label{sec:dm}}
A major open question for modern physics is the nature of the dark
matter (DM). There is a large body of evidence for the presence of
an unknown form of gravitational mass, at scales from kiloparsecs to
megaparsecs, that cannot be accounted for by SM particles. The
observation by the WMAP satellite~\cite{Komatsu:2010fb} of the
acoustic oscillations imprinted in the cosmic microwave background
quantifies the DM component as contributing about 25\% of the
total energy budget of the Universe.
Being dominant with respect to 
the baryonic component, which accounts for only about 4\% of the total
energy density, DM shaped the formation of cosmic structures. By
comparing the galaxy distributions in large redshift galaxy
surveys~\cite{Reid:2009xm}, and through $N-$body simulations of
structure formation~\cite{Springel:2008b,Anderson:2010df,Springel:2008by}, it is
inferred that the particles constituting the cosmological DM had to be
moving non-relativistically at decoupling from thermal equilibrium in
the early universe (`freeze-out'), in order to reproduce the observed
large-scale structure in the Universe and hence the term ``cold DM''
(CDM). This observational evidence has led to the establishment of a
concordance cosmological model, dubbed $\Lambda$CDM~\cite{Press:1974,
Sheth:1999su,  Springel:2005nw}, although this paradigm is troubled by
some experimental controversies~\cite{Klypin:1999uc,Kravtsov:2009gi,deNaray:2011hy,Walker:2011zu,BoylanKolchin:2011de,BoylanKolchin:2011dk}. 

One of the most popular scenarios for CDM is that of weakly
interacting massive particles (WIMPs), which includes a large class of
non-baryonic candidates with mass typically between a few tens of GeV
and few TeV and an annihilation cross-section set by weak interactions
\cite[see, e.g., Refs.][]{Bertone:2005a,Feng:2010gw}. Natural WIMP candidates are
found in proposed extensions of the SM, e.g. in Super-Symmetry
(SUSY)~\citep{Jungman:1996a, Martin:1998a}, but also Little
Higgs~\citep{Schmaltz:2005ky}, 
Universal Extra Dimensions~\citep{Servant:2003a}, and Technicolor
models~\citep{Nussinov:1985xr,Chivukula:1989qb}, among others.   
Their present velocities are
set by the gravitational potential in the Galactic halo at about a
thousandth of the speed of light. WIMPs which were in thermal
equilibrium in the early Universe would have a relic abundance varying
inversely as their velocity-weighted annihilation cross-section (for
pure $s-$wave annihilation): 
$\Omega_{\rm
  CDM}h^2=3\times10^{-27}\rm{cm}^3\rm{s}^{-1}/\left(\sigma_{\rm
  ann}v \right)$~\cite{Jungman:1996a}. Hence for a weak-scale
cross-section $\left(\sigma_{\rm ann}v\right) =
3\times10^{-26}\rm{cm}^3\rm{s}^{-1}$, they naturally have the required
relic density $\Omega_{\rm CDM}h^2=0.113\pm0.004$, where
$h=0.704\pm0.014$ is the Hubble parameter in units of $100$ km
s$^{-1}$ Mpc$^{-1}$~\cite{Komatsu:2010fb}. The ability of WIMPs to
naturally yield the DM density from readily computed thermal
processes in the early Universe without much fine tuning is sometimes
termed the ``WIMP miracle''.


In some SUSY theories, a symmetry called
`$R$-parity' prevents a too rapid proton-decay, and as a side-effect, also
guarantees the stability of the lightest SUSY particle (LSP), which is
thus a prime candidate for a WIMP. WIMPs can annihilate to SM
particles, and have hadron or leptons in the final products of
annihilation. Thus from cosmic DM annihilations, one can expect
emission of neutrinos, charged cosmic rays, multi-frequency
electromagnetic radiation from charged products, and prompt
gamma-rays~\cite{Colafrancesco:2005ji}.  The detection of these final
state particles can help to identify DM --- this is termed ``indirect
DM detection''. Gamma-rays are not deflected by cosmic
magnetic fields, and thus trace back to their origin. Therefore,
observation of a gamma-ray signal from cosmic targets where DM is
expected could prove conclusive about its nature . \\

In the context of gamma-ray astronomy, the differential flux of
gamma-rays from within a solid angle $\Delta\Omega$ around a given
astronomical target where DM is expected, can be written as:
\begin{equation}
\label{eqnp}
\frac{\mathrm{d}\Phi(\Delta\Omega,E_{\gamma})}{\mathrm{d}E_{\gamma}}\,= 
{\rm B_F}\cdot\frac{1}{4\pi}\,\underbrace{\frac{\left(\sigma_{\rm ann}
v\right)}{2\,m^2_\chi}\,\sum_i{\rm BR}_i\frac{\mathrm{d}N^i_{\gamma}}{\mathrm{d}E_{\gamma}}}_{Particle\,Physics}\,\cdot\,\underbrace{\widetilde{J}(\Delta\Omega)}_{Astrophysics}\, ,
\end{equation}
where $\left( \sigma_{\rm ann} v\right)$ is the 
annihilation cross-section (times the relative velocity of the two WIMPs), $\sum_i{\rm 
  BR}_i\,\mathrm{d}N^i_{\gamma}/\mathrm{d}E_{\gamma} =
\mathrm{d}N_{\gamma}/\mathrm{d}E_{\gamma}$ is the photon 
flux per annihilation summed over all the possible annihilation
channels $i$ with branching ratios ${\rm BR}_i$, and $m_\chi$ is the
mass of the DM particle. The `astrophysical factor' $\widetilde{J}$
is the integral over the line of sight (los) of the squared DM density
and over the integration solid angle $\Delta\Omega$:
\begin{equation}
\label{eqn:jbar}
\widetilde{J} = \int_{\Delta\Omega}d\Omega 
 \int_{\rm los}\mathrm{d}s\,\rho^2(s,\Omega).
\end{equation}
The remaining term ${\rm B_F}$ in Eq.~(\ref{eqnp}) is the so-called `boost factor'
 which is a measure of our ignorance of intrinsic flux
contributions that are not accounted for directly in the
formula. 

There are various known mechanisms for boosting the intrinsic
flux, among which we mention the inclusion of subhalos, and the
existence of a `Sommerfeld enhancement' of the cross-section at low
velocity regimes in models where the DM particles interact via a new
long-range force. 
All numerical $N-$body simulations of galactic halos have shown the
presence of subhalos populating the host halo~\citep[see, e.g.,
  Refs.][]{Springel:2008b, Diemand:2008a}. Such density enhancements,
if not spatially resolved, can contribute substantially to the
expected gamma-ray flux from a given object. This effect is strongly
dependent on the target: in dwarf spheroidal galaxies (dSphs) for
example the boost factor is only of ${\cal O}(1)$~\citep{Pieri:2009,
  Abramowski:2010zzt}, whereas in galaxy clusters the boost can be
spectacular, by up to a factor of several
hundreds~\cite{SanchezConde:2011ap, Pinzke:2011ek, Gao:2011rf}.  On the
other hand, the Sommerfeld enhancement effect can
significantly boost the DM annihilation
cross-section~\citep{Sommerfeld:1932,Lattanzi:2009a}. 
This non-relativistic effect arises when two DM particles interact in
a long-range attractive potential, and results in a boost in gamma-ray
flux which increases with decreasing relative velocity down to a
saturation point which depends on the DM and mediator particle mass.
This effect can enhance the annihilation
cross-section by a few orders of magnitude~\citep{Pieri:2009,
  Abramowski:2010zzt}.


The current generation of IACTs is actively searching for WIMP
annihilation signals. dSphs are promising targets
for DM annihilation detection being among the most DM dominated
objects known and free from astrophysical background.
Constraints on WIMP annihilation signals from
dSphs have been reported towards Sagittarius, Canis Major, Sculptor
and Carina by H.E.S.S.~\citep{Aharonian:2007km, Aharonian:2008dm,
  Abramowski:2010zzt}, towards Draco, Willman 1 and Segue 1 by
MAGIC~\citep{Albert:2007xg, Aliu:2008ny, Aleksic:2011jx}, towards
Draco, Ursa Minor, Bo\"{o}tes 1, Willman 1 and Segue~1 by
VERITAS~\citep{Acciari:2010a, Aliu:2012ga}, and again towards Draco and Ursa
Minor by Whipple~\citep{Wood:2008a}.  Nevertheless, the present
instruments do not have the required sensitivity to reach the
``thermal'' value of the annihilation cross-section $\left(\sigma_{\rm ann} v\right)
= 3\times10^{-26}\rm{cm}^3\rm{s}^{-1}$.
A search for a WIMP annihilation signal from the halo at angular
distances between 0.3$^{\circ}$ and 1.0$^{\circ}$ from the Galactic
Centre has also recently been performed using 112\,h of H.E.S.S.~data
\cite{Abramowski:2011}. For WIMP masses well above the H.E.S.S.~energy
threshold of 100\,GeV, this analysis provides the currently most
constraining limits on $\left(\sigma_{\rm ann} v\right)$ at the level of a few$\times 10^{-25}\,\mbox{cm}^3\mbox{s}^{-1}$.
H.E.S.S., MAGIC and VERITAS have also observed some galaxy clusters,
reporting detection of individual galaxies in the cluster, but only
upper limits on any CR and DM associated
emission~\cite{Aharonian:2008uq, Aharonian:2009, Aleksic:2009ir,
  Acciari:2009uq, Aleksic:2011cp, Abramowski:2012}.  Even though IACT 
limits are weaker than those obtained from the Fermi-LAT satellite
measurements in the GeV mass range~\citep{Abdo:2010b,
  Abazajian:2010sq, Hutsi:2010ai, Ackermann:2011},
they complement the latter in the TeV mass range.  Gamma-ray line
signatures can also be expected in the annihilation or decay of DM
particles in space, e.g. into $\gamma\gamma$ or $Z^0\gamma$. Such a
signal would be readily distinguishable from astrophysical gamma-ray
sources which typically produce continuous
spectra~\cite{Bringmann:2011ye}. A measurement 
carried out by H.E.S.S.~\cite{Spengler:2011lines} using over 100~h of
Galactic Centre observations and over 1000~h of extragalactic
observations complements recent results obtained by
Fermi-LAT~\cite{Abdo:2010nc}, and together cover about 3 orders of
magnitude in energy, from 10 GeV to 10~TeV.\\
%
%

In this contribution, we focus on the prospects for DM
searches with CTA,
which are expected to improve on the current
generation of IACTs on the following basis: 
\begin{list}{--}{\itemsep=0pt}
\item the energy range will be extended, from a few tens of GeV to
  several tens of TeV. At low energies, this will allow overlap with
  the Fermi-LAT instrument, and will provide sensitivity to WIMPs with
  low masses. For WIMPs with mass larger than about 100 GeV, CTA will
  have higher sensitivity as our studies indicate \citep{Funk:2012}.
\item the improved sensitivity in the entire energy range, compared to
  current instruments, will obviously improve the probability of
  detection, or even \emph{identification} of DM, through the
  observation of spectral features,
\item the increased FOV (about 10 deg versus $2-5$ deg)
  with a much more homogeneous sensitivity, as well as the improved
  angular resolution, will allow for much more efficient searches for
  extended sources like galaxy clusters (Section~\ref{sec:gc}) and
  spatial anisotropies (Section~\ref{sec:spatial}),
\item finally, the improved energy resolution will allow much better
  sensitivity to the possible spectral feature in the DM-generated
  photon spectrum. While astrophysical sources show typically
  power-law spectra with steepening at high energies, DM
  spectra are universal and generically exhibit a rapid cut-off at the
  DM mass. For specific models, ``smoking gun'' spectral features can
  appear \citep{Bringmann:2011ye}. The observation of a few identical
  such spectra from different sources will allow both precision
  determination of the mass of the WIMP and its annihilation
  cross-section.
\end{list}
 


For the following studies, in order to have a detection, we require
$a)$ the number of 
excess events over the background larger than 10 in the signal region,
$b)$ the ratio between the number of excess events and the background
events larger than 3\%, and $c)$ the significance of the detection
computed following Eq.~(17) of Li\&Ma \citep{Li:1983fv}, to be larger than
$5$. If not explicitly mentioned, we used a number of
background-control regions set to 5 ($\alpha$ = 0.2 in the Li\&Ma
notation), which is a conservative choice, given the fact that the
large FOV of CTA may allow for $\alpha<0.2$. In case of non detection
within a certain observation 
time, we calculate integral upper limits following the methods
described in \citet{Rolke:2005a} (bounded profile likelihood ratio
statistic with Gaussian background, and with a
confidence level of 95\% C.L) in all cases expect the Galactic halo
case, where we use the method of \citet{FeldmanCousins1998}.

We study the effect of various annihilation spectra, assuming in turn
100\% BR into a specific channel ($b\bar{b}$, $\tau^+\tau^-$ or
$\mu^+\mu^-$). The spectral shapes are obtained from
different parameterisation from the literature 
\citep{Tasitsiomi2002,Cirelli:2010xx,Cembranos2011}. For the
$b\bar{b}$ channel, which is used for comparison of different targets
(see Fig.~\ref{fig:Fermi.vs.CTA}), this difference accounts for few
percents (depending on the DM mass), which is substantially 
smaller than the uncertainties in, e.g.,  the astrophysical factor,
and do not significantly alters the conclusions.


\subsection{Observations of dwarf satellite galaxies\label{sec:dwarf}}
In the $\Lambda$CDM paradigm, galaxies such as ours are the result of
a complex merger history and are expected to have extended halos of DM
in accordance with observations. dSphs are satellites
orbiting the Milky Way under its gravitational influence and are
considered as privileged targets for DM searches for the following
reasons:
\begin{list}{--}{\itemsep=0pt}
\item the study of stellar dynamics shows that dSphs are among the
  most DM-dominated systems in the Universe, with mass-to-light ratio
  up to a few hundreds. In particular, the otherwise very uncertain
  astrophysical factor (Eq.~\ref{eqn:jbar}) can be constrained by dynamical
  arguments \citep{Evans:2003sc},
\item many of the dSphs lie within $\sim 100$ kpc  of the Earth,
\item they have favourable low gamma-ray backgrounds due to the lack
  of recent star formation history and little or no gas to serve as
  target material for cosmic-rays~\citep{Mateo:1998wg}.
\end{list}


The family of dSphs is divided into ``classical'' dSphs, which are
well-established sources with relatively high surface brightness and
hundreds of member stars identified~\cite{Simon:2007dq, Charbonnier:2011ft}, and 
``ultra-faint'' dSphs, which have mainly been discovered recently
through photometric observations in the Sloan Digital Sky Survey
(SDSS)~\cite{York:2000gk} and have very low surface brightness and
only a few tens or hundreds of member stars. Some of the ultra-faint
dSphs are not well-established as such because of similarity of their
properties with globular clusters, hence their nature is often under
debate. However, they are of particular interest due to their
potentially very
large, albeit uncertain, mass-to-light ratios.


\begin{table}[!t]
\centering
\small{%
\begin{tabular}{l|ccc|l|c}
\hline
dSph & Dec.  & D     & $\widetilde{J}$         & Profile & Ref.\\
    & [deg] & [kpc] & [GeV$^{2}$\,cm$^{-5}$] &  \\
\hline
Ursa Minor & $+44.8$ & $66$ & $2.2\times10^{18}$ & NFW & \cite{Charbonnier:2011ft}\\
Draco      & $+34.7$ & $87$ & $7.1\times10^{17}$ & NFW & \cite{Charbonnier:2011ft}\\
Sculptor   & $-83.2$ & $79$ & $8.9\times10^{17}$ & NFW & \cite{Charbonnier:2011ft}\\
           &         &      & $2.7\times10^{17}$ & ISO & \cite{Battaglia:2008jz}\\
Carina     & $-22,2$ & $101$& $2.8\times10^{17}$ & NFW & \cite{Charbonnier:2011ft}\\
\hline 
Segue~1    & $+16.1$ & $23$ & $1.7\times10^{19}$ & Einasto & \cite{Aleksic:2011jx}\\
Willman~1  & $+51.1$ & $38$ & $8.4\times10^{18}$  & NFW & \cite{Acciari:2010a}\\ 
Coma Berenices & $+23.6$ & 44 & $3.9\times10^{18}$ & NFW &
\cite{Strigari:2008} \\

\hline
\end{tabular}
}
\caption{\label{tab:jbar} Astrophysical factors for a selection of the
  most promising classical and ultra-faint dSphs. Dec. is
the target declination, D the distance and $\widetilde{J}$ is defined
as in Eq.~\ref{eqn:jbar}.}
\end{table}

Table~\ref{tab:jbar} shows the astrophysical factor $\widetilde{J}$
for few selected dSphs for comparison. For the classical dSphs, we selected the two
most promising Northern (Ursa Minor and Draco) and Southern (Sculptor
and Carina) ones according to~\citet[Table~2]{Charbonnier:2011ft}. The statistical
uncertainties on the 
astrophisical factor are roughly one order of magntiude at 68\% CL, 
slightly depending on the dSphs, and can be found in \citep[Table~2]{Charbonnier:2011ft}. 
For the ultra-faint dSphs, we
include Segue~1, Willman~1 and Coma Berenices, which have the highest
$\widetilde{J}$-values (although their nature is still under debate, especially for
Segue~1~\cite{Belokurov:2007a, NiedersteOstholt:2009a, Geha:2009a,
  XiangGruess:2009a, Simon:2010a, Essig:2009a, Martinez:2009jh}, which
makes the determination of the astrophysical factor less
accurate than for classical dSphs). 
We remark how the estimation of the astrophysical factor is subject to
uncertainties of either statistical origin or due to the different
assumptions considered for its calculation. A systematic study
has been done for Sculptor, to estimate the effect of the
profile shape and velocity anisotropy assumptions \cite{Battaglia:2008jz}. Another compilation of
astrophysical factors for several dSphs can be found in
\cite{Ackermann:2011}. 

For the subsequent discussion, we consider only three sources: Ursa Minor
and Sculptor representative of classic dSphs and located in the
Northern and Southern hemisphere respectively, and Segue~1 having the largest
astrophysical factor.

\subsection*{Bounds on the annihilation cross-section}
\begin{figure}[h!t]
\includegraphics[width=0.95\linewidth]{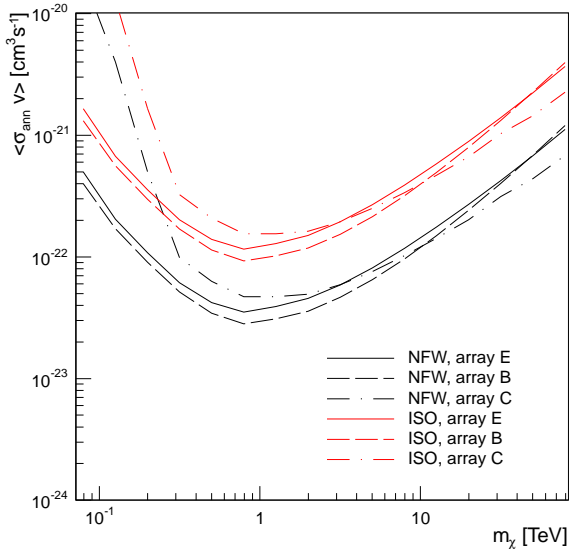}
\caption{\label{fig:profiles} CTA sensitivities on the
  velocity-averaged annihilation cross-section as a function of the
  WIMP mass for 100~hours
  observation of Sculptor with the CTA array $E$ (solid
  line), $B$ (dashed line) and $C$ (dashed-dotted line). Both the NFW (black line) 
  and cored isothermal (ISO, red line) DM halo profiles are shown, for
  an integration solid angle $\rm \Delta\Omega = 1\times10^{-5}\,sr$. 
  Annihilations are assumed to occur with 100\% branching ratio into
  $b\bar{b}$.}
\end{figure}

Two kinds of radial profiles are generally used to model the DM
distribution in dSphs: \emph{cusped} and \emph{cored}
profiles~\cite{Walker:2009zp}. While the former is
motivated by numerical $N-$body simulations, the latter seems to be more
consistent with observations~\cite{Salucci:2011ee}, but the issue is
still under debate ~\citep[see, e.g.,][]{Valenzuela:2005dh}. The
standard cusped profile is the Navarro, Frenk \& White form
(NFW)~\cite{Navarro:1997a}, while more recently it has been shown that
the Einasto profile~\cite{Navarro:2010a} provides also a good fit to
the subhalos in $N-$body simulations~\citep{Springel:2008b}. On
the other hand, for systems of the size of dSphs, the possibility of
centrally cored profiles has also been suggested~\citep{Moore:1994yx,
  Flores:1994gz,Walker:2011zu}. In conclusion, observations of low surface brightness
and dSphs~\citep{deBlok:2001fe,vandenBosch:2000rza,deBlok:2009sp} show
that both cusped and cored profiles can accommodate their stellar
dynamics.

Fig.~\ref{fig:profiles} shows the integral upper limits towards
Sculptor, the best Southern candidate from Table~\ref{tab:jbar}, for
which we consider both a cusped NFW~\cite{Charbonnier:2011ft} and a
cored isothermal~\cite{Abramowski:2010zzt} profile. The sensitivity is
calculated assuming that the DM particle annihilates purely in the
$b\bar{b}$ channel, for arrays $B,\;C$ and $E$. The observation time is set to 100 hours and
the integration solid angle to $\Delta\Omega = 10^{-5}$~sr.  The best
reached sensitivity is at the order of few$\times10^{-23}$~cm$^3$~s$^{-1}$ for
the NFW profile for both arrays $E$ and $B$, while the isothermal
profile is less constraining. Weaker constraints in the low mass
range are obtained for the $C$ array due to the lack of the large-size
telescopes in the centre of their layout. The capability of CTA to
discriminate between the two profiles is therefore restricted.

The integration solid angle plays a central role in the estimation of
the sensitivity and in the discrimination of the cusp or core
profiles. The former point was addressed already
\cite[Fig.~7]{Charbonnier:2011ft} where it was shown that small
integration angles guarantee the strongest constraints. In the case of
CTA, depending on the array layout (and the energy range), the
angular resolution could be as low as 0.02~deg, corresponding to a
minimum integration angle of about $10^{-6}$~sr, and thus our results
can be considered conservative, with an expected improvement of up to
a factor $\sim2$. Concerning the second point, \citet{Walker:2011fs}
showed that the more robust constraints, regardless of whether the
profile is cored or cusped, are reached for an integration angle
$r_c=2\,r_{1/2}/D$, where $r_{1/2}$ is the so-called half-light
radius, and $D$ is the distance to the dSph. For Sculptor, $r_c=0.52$,
which is over 5 times the integration angle adopted here. In our
calculation this would imply a weakening of the upper limits of a
factor of a few.\\

\begin{figure}[h!t]
\includegraphics[width=0.95\linewidth]{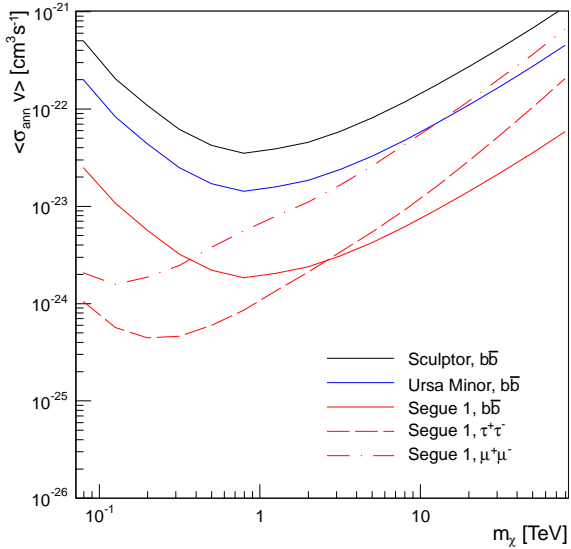}
\caption{\label{fig:sigmav_ul} CTA sensitivities on the
  velocity-averaged annihilation cross-section versus the WIMP mass for 100~hours
  observation towards Sculptor, Ursa Minor and Segue~1, assuming
  100\% branching ratio into $b\bar{b}$ (for Segue~1 also into
  $\tau^+\tau^-$ and $\mu^+\mu^-$). The calculations are done for
  array $E$ and $\rm \Delta\Omega = 1\times10^{-5}\,sr$. }
\end{figure}

In Fig.~\ref{fig:sigmav_ul} we show the integral upper limits 
for two classical dSphs, namely Ursa Minor and Sculptor in the
Northern and Southern hemispheres respectively, as well as the
ultra-faint dSph Segue~1. In order to span the variety of DM particle
models, we study the effect of various annihilation spectra (computed
using Ref.~\cite{Cirelli:2010xx}), assuming in turn 100\% BR into
$b\bar{b}$, $\tau^+\tau^-$ and $\mu^+\mu^-$ channels for the array $E$
and an observation time $\rm T_{\rm obs} = 100$~h. Assuming the
annihilation to be purely into $\tau^+\tau^-$, the sensitivity reaches
few $\times10^{-25}$~cm$^3$~s$^{-1}$ for 100~h
observation time of Segue~1. In comparing the different dSphs (assuming
the reference annihilation channel $b\bar{b}$) we see that even the
most promising classical dSphs are less constraining than Segue~1 by
over a factor of 10. However the uncertainties in the estimation of
astrophysical factors for ultra-faint dSphs mean that this conclusion
may not be reliable. Note that in the above calculations we did not
assume any intrinsic flux boost factor, i.e. ${\rm B_F}=1$ in
Eq.~(\ref{eqnp}).

\subsection*{Bounds on Astrophysical factors and Boost factors}
\begin{figure}[h!t]
  \includegraphics[width=0.95\linewidth]{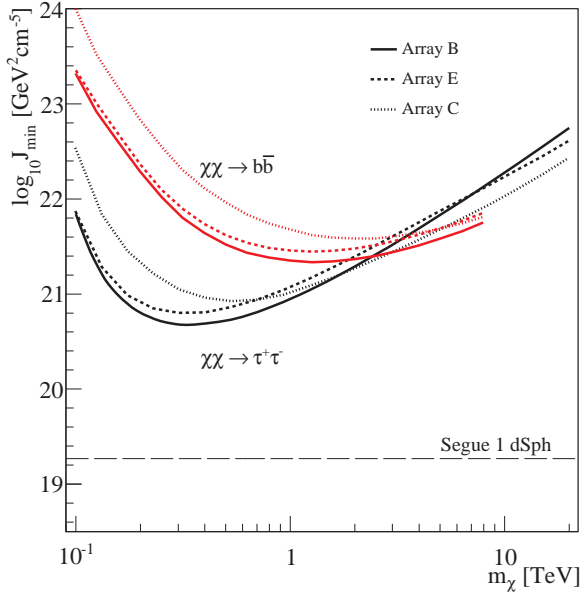}
  \caption{ \label{jfactors} The minimum value of the astrophysical
    factor required for a $5\,\sigma$ detection with $\rm T_{\rm obs}
    = 100\,h$, versus WIMP mass. Two annihilation channels are
    considered for arrays $B,$ $C,$ and $E$: $b\bar{b}$ (upper curves) and
    $\tau^+\tau^-$ (lower curves). The estimated astrophysical factor
    for Segue~1 is shown for comparison.}
\end{figure}

Another approach to estimate the capabilities of CTA for DM
detection in dSphs consists in the evaluation of the statistical
significance of the DM signal as a function of the DM particle mass
$m_{\chi}$ and the astrophysical factor, for different possible
annihilation channels.  Hereafter, we calculate the minimum
astrophysical factor $J_{\rm min}$ required to reach a statistical
significance of $5\,\sigma$ assuming an effective observation time of
$100$~h, and the thermal cross-section $3 \times 10^{-26}
\mbox{cm}^{3}\mbox{s}^{-1}$.  This is shown in Fig.~\ref{jfactors} for
two annihilation 
channels: $b\bar{b}$ (upper curves) and $~\tau^+\tau^-$ (lower
curves), using analytical fits from Ref.~\cite{Cembranos:2010xd}.  Again, three proposed CTA
configurations are studied: B, C, and E. In order to put these values
into context, we note that the largest astrophysical factor
$\widetilde{J}$ for known
dSphs is that of Segue~1 at $1.7\times
10^{19}$~GeV$^{2}$~cm$^{-5}$~\cite{Essig:2010em}. From the figure we
see that array $B$ is the most constraining over the whole energy range. It is
clear that for a detection, the astrophysical factor of the dSph needs
to exceed $10^{21}$~GeV$^{2}$~cm$^{-5}$, which is only 1--2 orders of
magnitude smaller than that of the Galactic Centre (see Section~\ref{sec:halo}). While we may
expect a few such objects in the Milky Way halo~\cite{Pieri:2008a},
they ought to have already been detected and identified by
Fermi-LAT. Although this has not happened, one can envisage DM subhalos
with no associated dSph (or one not bright enough optically to be
detected), and therefore such gamma-ray emitters may be hidden among
the unidentified Fermi sources~\cite{Nieto:2011sx}.\\

\begin{figure}[h!t]
\includegraphics[width=0.95\linewidth]{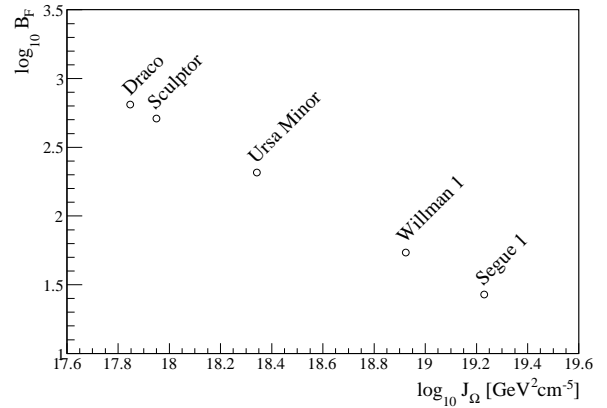}
\caption{\label{fig:Boost} Minimum boost factor required for a $5\,\sigma$
  detection in 100~h by array $B$, for the dSphs in
  Table~\ref{tab:jbar} and a 1~TeV WIMP annihilating into
  $\tau^{+}\tau^{-}$. The density profiles are taken to be NFW,
  except for Segue~1 where an Einasto profile has been
  assumed. The smallest boost required is ${\rm B_F}=25$ for Segue~1.}
\end{figure}

Another way to evaluate the prospects of DM detection is by means of
the intrinsic flux \emph{boost factor} term ${\rm B_F}$ in
Eq.~(\ref{eqnp}).  The
minimum ${\rm B_F}$ is computed as the ratio of the minimum
astrophysical factor $J_{\rm min}$ which provides a 5$\,\sigma$
detection in 100~h of observation time with CTA, to the observational
astrophysical factor $\widetilde{J}$ from the DM modeling of the
dSphs. Again, the thermal cross-section $3 \times
10^{-26} \mbox{cm}^{3}\mbox{s}^{-1}$ is assumed. Fig.~\ref{fig:Boost} shows the minimum ${\rm B_F}$ for a 1~TeV
DM particle annihilating into $\tau^{+}\tau^{-}$. $J_{\rm obs}$ is
calculated for a NFW profile for all the cases except Segue~1, where an
Einasto profile is considered. Considering that the
boost factor from subhalos in dSph is only of ${\cal O}(1)$, CTA
observations of dSphs will be more sensitive to scenarios where 
Sommerfeld enhancement is at play, which may instead boost the signal
up to ${\cal O}(1000)$.

%
%
%
\subsection{Observations of Galaxy Clusters\label{sec:gc}}

Within the standard $\Lambda$CDM scenario, 
galaxy clusters, with masses around $10^{14}-10^{15}$~M$_\odot$, are
the largest gravitationally bound objects and the most recent structures to
form~\cite{Voit:2004ah}. They are complex objects, relevant for both
cosmological and astrophysical studies, and for what concerns DM
searches \cite{Colafrancesco:2005ji, SanchezConde:2011ap,
  Pinzke:2011ek, Blasi:2007pm,Jeltema:2008vu,
  Pinzke:2010st,Ackermann:2010rg, Cuesta:2011, Zimmer:2011vy, Huang:2011dq, Han:2012au}. DM, in fact, is supposed to be the
dominant component of the cluster mass budget, accounting for up to 80\%
of its mass (the other components are the galaxies and the gas of the
intra-cluster medium (ICM)). This is why clusters have been considered
as targets for the indirect detection of DM, with the possibility of
detecting the gamma-rays produced in the annihilation (or decay) of DM
particles in the halo of the cluster. 

$N-$body simulations of halo formation and evolution have also proven
that, while the majority of early-formed, small structures merge
together giving shape to more massive objects, some of the
subhalos survive and are still present in the ``host'' halo of
larger objects. Theoretical models foresee a huge number of these
substructures at all scales down to $10^{-11}-10^{-3}$~M$_\odot$
\cite{Bringmann:2009vf}.  
These subhalos have the effect of contributing to the total
gamma-ray emission from DM annihilations, and they may have important
consequences for DM indirect detection. This is especially true for
galaxy clusters, where the intrinsic flux ``boost'' from subhalos
can be of order $100-1000$, in particular compared to the case of
dSphs, explored previously, where the subhalos boost should
contribute only marginally. Despite the fact that, due to their vicinity, dSphs
are usually considered as the best sources for DM indirect detection,
thanks to the subhalos boost, some authors claim that galaxy
clusters have prospects of DM detection better or at least as good as
those of dSphs \cite{SanchezConde:2011ap,Pinzke:2011ek, Gao:2011rf}.

On the other hand, in galaxy clusters, emission in the gamma-ray range
is not only expected by DM annihilation. Clusters may host an Active
Galaxy Nucleus (AGN, that appear as point-like sources at very high
energies) and radio galaxies.  The
case of the Perseus galaxy cluster, which has been observed by MAGIC
during several campaigns in the last years, is emblematic: MAGIC
detected both the central AGN NGC-1275~\cite{Aleksic:Perseus} and the
off-centreed head-tail radio galaxy
IC~310~\cite{Aleksic:2010xk}. Moreover gamma-rays are expected to be
produced also from the interaction of cosmic rays (CRs) with the ICM
\cite{Pinzke:2010st, Blasi:1999aj,Miniati:2003ep,Pfrommer:2007sz,
  Ensslin:2010zh}. The physics of the acceleration of CRs
(electrons and protons) is not completely understood, but plausible
mechanisms can be shock acceleration during structure formation, or galactic
winds driven by supernovae. CRs can also be injected into the ICM from
radio galaxy jets/lobes. At the energies of interest here (above 10
GeV), CRs emit gamma-rays from the processes associated with the decay
of the neutral and charged pions produced in the interaction of the
CRs with the ICM ambient protons
\cite{Colafrancesco:1998us,Colafrancesco:2010kx}. Most importantly,
such a contribution is usually found to be larger than the one
predicted from DM annihilation. It thus represents an unavoidable
source of background for DM searches in galaxy clusters. To date, the deep exposure
performed with the MAGIC stereoscopic system of the Perseus
cluster~\cite{Aleksic:2011cp} placed the most stringent constraints
from VHE gamma-rays observations regarding the maximum CRs-to-thermal pressure to
$\langle X_{CR}\rangle<1-2$\%.  

The purpose of this section is to estimate the CTA potential to detect
gamma-rays from DM annihilation in the halo of galaxy clusters. First,
the CR-induced emission only will be considered. This component
represents, by itself, an extremely interesting scientific case, at
the same time being a background complicating the prospects of DM
detection.  Afterward, the ideal case of a cluster whose
emission is dominated by DM annihilation only will be treated.
Finally, the combination of the two components distributed
co-spatially will be discussed.

It should be noted here that gamma-ray emission from both DM
annihilation and CRs is spatially extended, even though not always
co-spatial. In particular, \citet{SanchezConde:2011ap} proved that,
for the case of DM, the contribution of subhalos is particularly
relevant away from the halo centre, so that annihilations can
still produce a significant amount of photons up to a distance of
$1-2$ degrees from the centre.  This represents a problem for current
Cherenkov Telescopes since their FOV is limited to
$3-5$ degrees. CTA will overcome this limitation, having a
FOV of up to 10~deg (at least above 1 TeV) and an almost flat
sensitivity up to several 
degrees from the centre. It is reasonable to expect,
therefore, that CTA will allow a step-change in capability in this
important area. 

In this study, we selected two benchmark galaxy clusters: Perseus and
Fornax. Perseus has been chosen because it is considered that with
the highest CR-induced photon yield but a low DM content, and Fornax for the
opposite reason: it is considered the most promising galaxy cluster
for DM searches~\cite{SanchezConde:2011ap, Pinzke:2011ek}. We recall
that Perseus is located in the Northern hemisphere, while Fornax is in
the Southern hemisphere.  To study the prospects for CTA we use two
Monte Carlo simulations of the instrument response functions and of
the background rates for \emph{extended sources}, for the case of
array $B$ and array $E$, which we recall, are representatives of
well-performing arrays at low energies (array $B$) and in the full
energy range (array $E$). The MC simulations were developed explicitly
for the analysis of extended sources so that all the relevant
observables are computed throughout the entire FOV.

\subsection*{Gamma-ray emission from cosmic-rays}
\label{sec:cr}

Gamma-ray emission due to the injection of CRs into the ICM of a
galaxy cluster is proportional both to the density of the ICM and the
density of CRs. For the present work, we refer to the hadronic CR model of Pinzke
and collaborators \cite{Pinzke:2011ek, Pinzke:2010st}, based on
detailed hydrodynamic, high-resolution simulations of the evolution of
galaxy clusters, since in these works we found detailed morphological
information, essential to compute the CTA response. The CR surface
brightness rapidly decreases with the distance from the centre of the
halo, so that, in most cases, the total emission is contained in
$0.2-0.3\,r_{200}$, where $r_{200}$ is the projected virial radius of
the cluster, where the local density equals 200 times the critical density (see, e.g., Fig.~14 of \citet{Pinzke:2011ek}, from which 
we derive the surface brightness of the clusters we
analyze). 
$r_{200}=1.9$~Mpc ($1.4^\circ$) for Perseus and $0.96$~Mpc
($2.8^\circ$) for Fornax~\citep{SanchezConde:2011ap}. 
The energy spectrum of the model, in the energies of interests
here (above 10~GeV), is a power-law with a slope of $-2.25$.

\begin{figure}[h!t]
\centering
\includegraphics[width=0.95\linewidth]{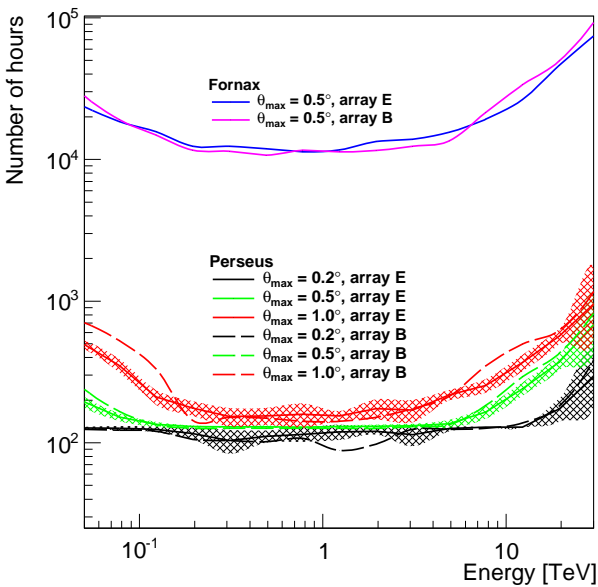}
\caption{\label{fig:Perseus_CR} Integration time required to have a
  5$\sigma$ detection (see text for details) of gamma-rays from
  CR-induced gamma-rays only according to the model
  of~\citet{Pinzke:2011ek} for Perseus (lower curves,
  $\Phi_{\mbox{\tiny{CR}}}(>100$~GeV$) = 2.04\times 10^{-11}
  \mbox{cm}^{-2}\mbox{s}^{-1}$) and Fornax (upper curves,
  $\Phi_{\mbox{\tiny{CR}}}(>100$~GeV$)=1.5\times 10^{-13}
  \mbox{cm}^{-2}\mbox{s}^{-1}$). The integration time is shown as a
  function of the energy threshold, and over different ROI, for the
  case of array $B$ (dashed lines) and array $E$ (solid lines). The
  shaded regions indicate the $1\sigma$ standard deviation among 10
  different simulations.}
\end{figure}

Since the emission region is extended in the sky, we first
divide the FOV into a grid of pixels each 0.2 degrees wide, and then
we define the region of interest (ROI), constituted by all the pixels
within an angle $\theta_{max}$ from the centre of the camera. We consider
15 values of energy threshold $E_i$ in logarithmic steps from 50 GeV
to 50 TeV. With the theoretical gamma-ray emission and the instrument
response, we are able to compute the predicted number of background
($N_i^{\mbox{\tiny{OFF}}}$) and signal events ($N_i$) above each
$E_i$, in each bin of the ROI separately, and then we integrate over
the entire ROI. 
The model of \citet{Pinzke:2011ek} predicts a rather large gamma-ray
flux for Perseus ($\Phi_{\mbox{\tiny{CR}}}(>100$~GeV$) = 2.04\times
10^{-11} \mbox{cm}^{-2}\mbox{s}^{-1}$), the largest among the galaxy
clusters, and a smaller one for Fornax
($\Phi_{\mbox{\tiny{CR}}}(>100$~GeV$)=1.5\times 10^{-13}
\mbox{cm}^{-2}\mbox{s}^{-1}$). Above the different energy thresholds $E_i$,
we determine how many hours CTA will need to detect the sources.  We
perform the calculation for the two CTA array $B$ and $E$ and for
different ROI. We repeat the procedure 10 times for each energy
threshold and average the results, in order to quantify the
statistical fluctuations occurring when the number of events (both
$N_i$ and $N_i^{\mbox{\tiny{OFF}}}$) are generated. The results are
shown in Fig.~\ref{fig:Perseus_CR}.

If one assumes the CR-induced gamma-ray model by
\citet{Pinzke:2011ek}, CTA will detect such radiation from Perseus
already in about $100$~h, a fact which will constitute an
extraordinary scientific result by itself\footnote{We underline that
  the upper limits obtained by the MAGIC experiment on 
Perseus~\citep[Fig.~3]{Aleksic:2011cp} already constrain by about 20\% the model
predictions (the same used here), impling that the maximum CR
acceleration efficiency is lower than 50\% or, alternatively, the
presence of non-neligible CR transport phenomena.}. The discovery could indeed be potentially
close, opening up a completely new observation window on the Universe. 
We underline that there is an absolute lower limit for gamma-rays in
the hadronic scenario for clusters with an observed radio halo: 
a stationary distribution of CR electrons loses all its energy to
synchrotron radiation for strong magnetic fields, as those in the radio
halo, and therefore the ratio of gamma-ray to synchrotron flux becomes
independent of the spatial distribution of the CRs and the thermal
gas. 
For
the Perseus cluster this lower limit is roughly a factor $3-4$ from
the gamma-ray flux predicted by the CR model~\cite[see Fig 3 in
  Ref.][]{Aleksic:2011cp}, hence CTA would, in the
worst case scenario, require about $1,000$ hours of observation to
completely rule out the hadronic models. Such large observation times can in
principle be achieved either by, e.g., multi-annual observational
campaigns. On the other hand, a non detection with CTA in a few hundred hours would seriously constrain the model and thus pose interesting
challenges on the galaxy cluster physics. 
The situation is more pessimistic for our model of Fornax, which is
out of reach for CTA.  

We see that the exact value of the integration time depends on the energy threshold
chosen for the analysis. The reason for this is the tradeoff between
the gamma-ray efficiency at different energies (the effective area),
the source intrinsic spectrum and the chosen ROI.
Roughly 90\% of the CR-induced emission is expected within about
$0.1\,r_{200}$ for Perseus, which corresponds to roughly $0.2^\circ$. We checked that integrating larger ROI, 
more background than signal is included in the analysis, thus
deteriorating the significance of the detection. This suggests that in
realistic cases, the best ROI should be optimized.
Finally, we also see that the prospects of detection are similar for
both considered arrays, $B$ and $E$.

\subsection*{Gamma-ray emission from Dark Matter annihilation}
The gamma-ray brightness due to DM annihilations from a particular
viewing angle in the sky is proportional to the DM
density squared integrated along the line of sight, as shown in
Eq.~(\ref{eqnp}). In the case of galaxy clusters, the contribution of
the smooth DM halo is boosted by the presence of DM
subhalos. Recent $N$-body simulations of Milky Way-like halos
\cite{Springel:2008b, Anderson:2010df, Springel:2008by} found that the
contribution of subhalos is small in the centre of the halo, due
to dynamical friction and tidal effects that disrupt the
subhalos. However, already at distances of $0.01-0.05\,r_{200}$,
subhalos become the dominant component. The real value of the boost factor from
subhalos is unknown and the theoretical estimates depend on
different assumptions and different methods used in the
calculations. \citet{Pinzke:2011ek} estimated a ${\rm B_F}=580\mbox{
  and }910$ for Fornax and Perseus respectively (for a minimal halo
mass of $10^{-6}$~M$_\odot$) , while other authors
gave ${\rm B_F}$ from few tens \cite{SanchezConde:2011ap} up to
several thousands \cite{Gao:2011rf}.

We refer again to the results of \citet{Pinzke:2011ek} where the
authors assumed a double power-law to describe the luminosity of
subhalos as a function of the projected distance from the centre
of the halo, a behavior derived by analyzing the sub-halos in the
Aquarius $N$-body simulation. They also found the projected surface
brightness to be largely independent of the initial profile of the
smooth DM halo. As a result, the DM profile is very
flat since the emission decreases approximately only $10$\% at a
distance of $1.5-2.0$ degrees from the centre, depending on the
cluster~\cite[Fig.~\ref{fig:morpho} and Ref.][]{SanchezConde:2011ap}. For the case of
Perseus and Fornax, we used the results of Fig.~10 of
Ref.~\cite{Pinzke:2011ek}, assuming a telescope angular resolution of
$0.1$ degree, which is a good approximation for CTA, despite the fact
that the exact value depends on the array, the energy and the position
in the FOV. We underline that in the case of galaxy cluster, the contribution from substructure strongly shapes the region of emission, basically moving from a point-like source (in case no substructure are considered), to an extended source. Given the fact that the analysis differ in the two cases, the contribution from substructure cannot be considered as a simple multiplicative factor in the intrinsic expected flux with respect to point-like case.

Hereafter we consider the Fornax cluster, which has the largest
expected DM-induced photon yield. The intrinsic flux is taken from
\cite[Table~2]{Pinzke:2011ek} and includes an intrinsic boost factor
from subhalos of ${\rm B_F}=580$, summing up to a total flux of
$\Phi_{\mbox{\tiny{DM}}}(>100$~GeV$)=3.6\times 10^{-13}
\mbox{cm}^{-2}\mbox{s}^{-1}$. Additional intrinsic boost factor may
come from either other contributions from subhalos not accounted in
this model, by mechanisms like the Sommerfeld enhancement discussed
above, or by the effect of contraction processes due to baryonic 
condensation~\cite{Gnedin:2004cx,Ando:2012vu}.

\begin{figure}[h!t]
\includegraphics[width=0.95\linewidth]{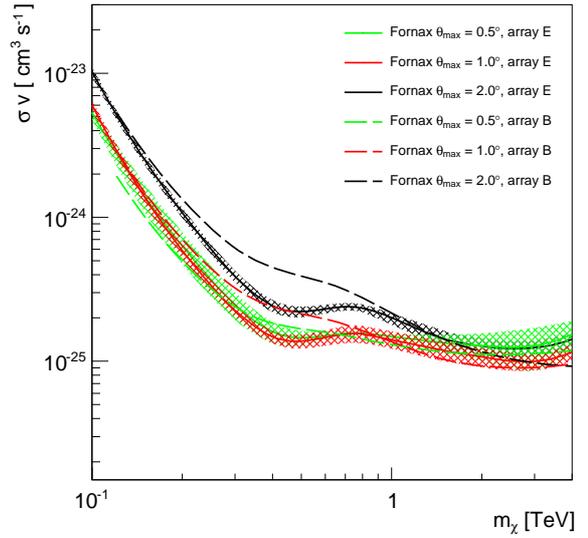}
\caption{\label{fig:Fornax_DM} Prospect of detection of DM-induced
  signal from Fornax for a DM particle annihilating into $b\bar{b}$ and
  100~h integration time. The reference model is taken from
  Ref.~\cite{Pinzke:2011ek} with subhalo boost factor ${\rm B_F}=580$.
  The   shaded regions indicate the $1\sigma$ standard deviation among 10
  different simulations.}
\end{figure}

To compute the CTA prospects of
detection, we consider only the case of DM annihilating into
$b\bar{b}$ (spectral shape obtained from Ref.~\citep{Cembranos:2010xd}), while other channels like $\tau^+\tau^-$ or $\mu^+\mu^-$
may be more constraining, depending on the energy (see Fig.~\ref{fig:sigmav_ul}). We take the
reference thermal cross-section $3 \times 10^{-26}
\mbox{cm}^{3}\mbox{s}^{-1}$ and we scan DM particle mass $m_\chi$
between 50~GeV and 4 TeV. We optimized the upper limit calculation as
described in Ref.~\cite{Aleksic:2011jx}, by optimizing the energy
threshold above which the upper limit is estimated. In addition, we consider the
possibility of extending the size of the ROI up to a $\theta_{max}$ of
2 degrees, to encompass the full radial extension of the source. 
%
Fig.~\ref{fig:Fornax_DM} show the results. In 100~h observation, the lack of
detection would place exclusion limits at the level of
$10^{-25}\rm{cm}^{3}\rm{s}^{-1}$. 

 We also
studied the effect of integrating over larger and larger regions:
despite the increased numbers of background events, the signal yield is
also larger and, in the case of Fornax, we gain more in integrating up
to $\theta_{max}=1^\circ$ than $0.5^\circ$, while integrating over
larger regions leads to a worse sensitivity.

\subsection*{Distinguishing the dark matter signal from other gamma-ray
  contributions}

\begin{figure}[h!t]
  \includegraphics[width=0.95\linewidth]{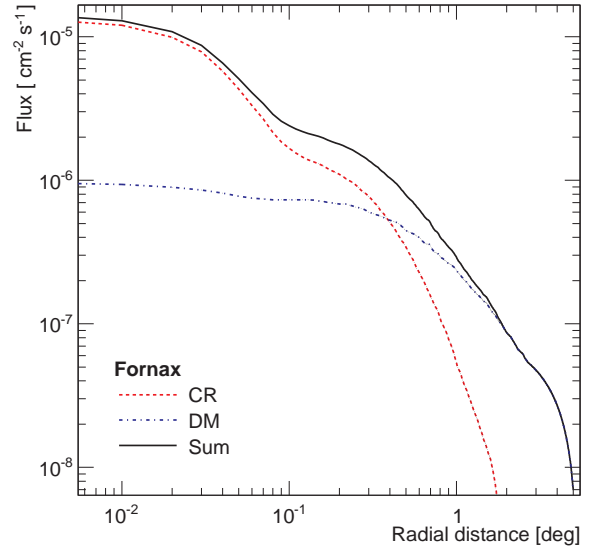}
  \caption{\label{fig:morpho} The surface brightness (above 1 GeV) of
    the gamma-ray emission from the Fornax cluste from CRs
    (red), DM (blue) and the sum of the two contributions (black). The
    DM emission is calculated from the K' benchmark model of
    \citep{Bringmann:2008kj} which has mass of 570~GeV and a
    velocity-averaged cross-section of $4.4\times10^{-26}$ cm$^3$
    s$^{-1}$. Adapted from Ref.~\citep{Pinzke:2011ek}. } 
\end{figure}

In the previous sections we have considered separately the
contributions of CR and DM to the total gamma-ray photon yield. This
is an unrealistic situation: galaxy clusters are, in fact, complex
objects where gamma-rays may be due to different contributions
possibly of different spatial origin: by collisions of accelerated CRs,
by DM annihilations \emph{and} by foreground or embedded astrophysical
sources.

Fortunately, gamma-rays of different origin typically have different
spectral shapes, with the DM-induced emission characterized by the
peculiar cut-off at $E=m_\chi$ and other remarkable spectral features
\cite{Bringmann:2007nk, Bringmann:2008kj}, in contrast to the
plain spectral shapes (typically power-laws within the energy range of
interest here) of the emission due to
CRs, of the central galaxy or 
any astrophysical objects in the cluster. In the 
case a VHE emission is detected from a cluster, this fact may be used
as a probe to discriminate between the components. However, we remark
that in order to significantly discriminate the two sources one would need
a quite significant detection over the CR-signal, which is often not
supported by theoretical predictions for most galaxy clusters.

A distinct approach could be based on the different spatial extensions
of the various contributions of VHE gamma-ray photons from galaxy
clusters. The possible individual galaxies emitting
within the cluster are typically seen as point-like sources, and thus one may
exclude them from the FOV for CR and DM searches.  Moreover,
from the fact that CR-induced radiation is more concentrated than that
induced by DM, one can optimize the ROI to
select only those where the emission is DM dominated.
In Fig.~\ref{fig:morpho}, we show the expected brightness profile for
CR and DM photons for the Fornax cluster. One can see that up to
$\theta=0.4^\circ$ the emission is dominated by CR-induced photons,
whereas this exact value is cluster-dependent and model-dependent, and
in particular the possible intrinsic boost-factor in the DM signal can
affect this. In this example, above $\theta=0.4^\circ$, the CR-signal fades more
rapidly than the DM one. Then, in principle, by considering a ROI with
a $\theta_{min}=0.4^\circ$, one could be able to isolate the DM
signal. The maximum integration angle $\theta_{max}$ should be
optimized according to the specific cluster and emission profile to
maximize the sensitivity, as discussed above. Unfortunately, at the
moment of writing this report, we did not have sufficient coverage in
the MC of extended sources to perform such a study, and we are limited
to a qualitative discussion. We mention that the ``geometrical''
discrimination makes sense only if the DM signal is sufficiently
large, otherwise different observational strategies could be more
constraining. Finally, we stress again that with a large FOV (at least
above 1 TeV) that has a near constant
sensitivity over several degrees will allow CTA to study extended
high energy gamma-ray sources in detail for the first time, with possibly
revolutionary consequences for the IACT technique.

\subsection{Observations of the Galactic Halo and Centre\label{sec:halo}}

The Galactic Centre (GC) is a long-discussed target for indirect DM searches with
Cherenkov telescopes~\citep{Gondolo:1999ef}. The density of the DM halo should be highest in
the very centre of the Milky Way, giving rise to a gamma-ray flux from
annihilation of DM particles. On the one hand, this view is
strengthened by the results of recent $N-$body simulations of CDM halos
\citep{Aquarius2008} suggesting that, for an observer within the  
Milky Way, the annihilation signal from DM is not primarily due to
small subhalos, but is dominated by the radiation produced 
by diffuse DM in the main halo. On the other hand, searches close to the 
GC are made difficult by the presence of the Galactic Centre source
HESS J1745$-$290 \citep{HessGC2010, HessGC2009} and of diffuse emission from the
Galactic plane \citep{HessRidge2006}. Both emissions can be plausibly
explained by astrophysical emission processes: HESS J1745$-$290 is
thought to be related to the Black Hole Sgr A$^{\ast}$ or the pulsar
wind nebula G 359.95$-$0.04 \citep{GCPWN2006}, and the diffuse emission
is well described as arising from hadronic cosmic rays interacting in
giant molecular clouds. In both cases, the measured energy spectra do
not fit DM model spectra \citep{HessGC2006} and thus make a dominant
contribution from DM annihilation or decay unlikely. 

In this situation, DM searches should better target regions which are
outside the Galactic plane and hence not polluted by astrophysical
gamma-ray emission, but which are still close enough to the GC 
to exhibit a sizable gamma-ray flux from DM annihilation in
the Milky Way halo~\citep{Schwanke2009}. Given the angular resolution
of Cherenkov telescopes and the scale height of the diffuse emission
from the  Galactic plane these criteria are fulfilled for an angular
distance of about $0.3^{\circ}$ from the GC. This angular
scale translates into a distance of 45\,pc from GC
when using 8.5\,kpc as the galactocentric distance. The radial DM density
profiles obtained in $N-$body simulations of Milky Way sized galaxies, like
Aquarius \citep{Aquarius2008} and Via Lactea II \citep{ViaLactea2008},
can be described by Einasto and NFW parameterizations,
respectively. These parameterizations differ substantially when
extrapolating to the very centre of the Milky Way halo since the NFW
profile is much more strongly peaked. At distances greater than about
10\,pc, the difference is, however, just a factor of 2 which implies
that a search at angular scales of $>0.3^{\circ}$ will not be hampered
by the imprecise knowledge of the DM density profile at small scales.

A search for a DM annihilation signal from the halo at angular
distances between 0.3$^{\circ}$ and 1.0$^{\circ}$ from the GC
has recently been performed using 112\,h of H.E.S.S.~data
\citep{Abramowski:2011}. For WIMP masses well above the H.E.S.S.~energy
threshold of 100\,GeV this analysis provides the currently most
constraining limits on the velocity averaged annihilation cross
section $\left(\sigma_{\rm ann} v\right)$ of WIMPs (for IACTs) at the level of few
$10^{-25}\,\mbox{cm}^3\mbox{s}^{-1}$. Towards lower WIMP masses, 
observations of dwarf galaxies with the Fermi-LAT satellite yield
even better limits \citep{Abdo:2010b} demonstrating how both
observations of dwarf galaxies and of the extended GC
region allow to jointly constrain the parameter space. 

\subsection*{Simulations and Assumptions}
The prospects of a search for DM annihilation photons from the Milky
Way halo with CTA depend on $(i)$ the performance of the southern CTA
array, $(ii)$ the applied analysis and background rejection techniques,
and $(iii)$ the details of the DM distribution and WIMP annihilation. At
low energies, the sensitivity of IACTs is limited by the presence of
hadron and electron showers which arrive isotropically and which can
only be distinguished from photons on a statistical basis. The basic
strategy for the halo analysis is therefore to compare the fluxes of
gamma-like events from a signal region (with solid angle $\Delta\Omega_s$)
and a  background region (solid angle $\Delta\Omega_b$) and to search for DM
features in the background-subtracted energy spectra. The signal
region can be chosen such  that it has the same instrumental
acceptance as the background region, but is located closer to the
GC and features therefore a higher DM annihilation
flux. For the purpose of this section, we rewrite Eq.~(\ref{eqnp}) in
terms of differential DM photon \emph{rate} expected from the signal or
background regions ($s,\,b$ respectively), given by:
%
\begin{equation}
 \frac{dR}{dE}|_{\,\mathrm{s,b}} = \frac{\left(\sigma_{\rm ann} v\right)}{8\pi\,m^2_\chi} \frac{dN_{\gamma}}{dE_{\gamma}}
   \int_{\Delta\Omega_{\,\mathrm{s,b}}} \negspcfive J(\Omega) A(\Omega,E) d\Omega\mbox{,}
\end{equation}
where $dN_{\gamma}/dE_{\gamma}$ is the photon spectrum generated in the annihilation of 
a WIMP of mass $m_\chi$, and $A(\Omega,E)$ are the CTA effective areas for
photons, which depend on the position of the region within the FOV ($\Omega$), the energy $E$ and further parameters (like the 
zenith angle of the observations). $J(\Omega)$ is the line-of-sight integral over the squared DM
density  $\rho(r)$ (cf.~Eq.~\ref{eqn:jbar}). Since the DM density depends only on the distance
to the GC $r$ the line-of-sight integral and the astrophysical factor are only a function
of the angular distance $\psi$ from the GC. Assuming that
the signal and background region differ only with respect to their DM
annihilation flux and their relative size $\alpha = \Delta\Omega_{\mathrm{s}}/\Delta\Omega_{\mathrm{b}}$, the rate of excess photon
events  $R_{\mathrm{s}} - \alpha R_{\mathrm{b}}$ is given by
\begin{equation}
    \frac{\left(\sigma_{\rm ann} v\right)}{8\pi m^2_\chi} \int_0^{\infty}\negspcfive dE\frac{dN_{\gamma}}{dE_{\gamma}} \left[
      \int_{\Delta\Omega_{\mathrm{s}}}\negspcfive J(\psi) A(\Omega,E) d\Omega - \alpha\int_{\Delta\Omega_{\mathrm{b}}}\negspcfive J(\psi) A(\Omega,E) d\Omega
    \right]
    \mbox{.}
  \label{eq:halo_rate}
\end{equation}

Clearly, the rate vanishes when the astrophysical factors
of the signal and the background regions are identical which implies that 
in the case of an isothermal DM density profile, a halo analysis with
signal and background region chosen too close to the GC will not
allow the placement of limits on $\left(\sigma_{\rm ann} v\right)$. 

\begin{figure}[h!t]
\centering
\includegraphics[width=0.75\linewidth]{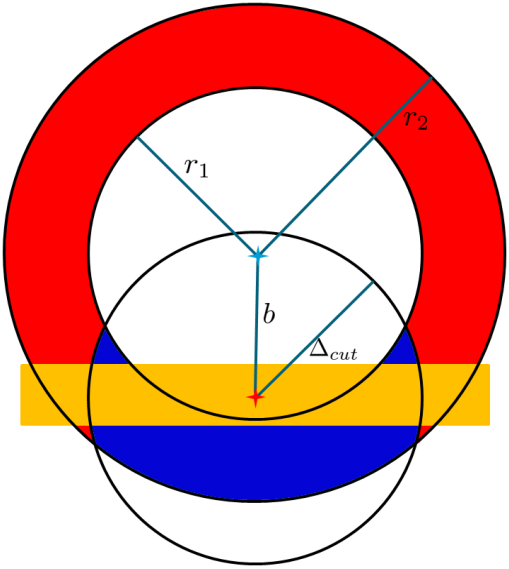}
\caption{\label{fig:halo_rings}Illustration of the {\em Ring Method}
  for constructing signal and background regions within one FOV of the CTA candidate arrays. The red star denotes the position
  of the GC in galactic coordinates; the blue star marks the pointing
  position of the CTA array which is shifted by an amount $b$ in
  latitude from the GC. The annulus with inner and outer radii $r_1$
  and $r_2$ around the observation position defines regions of equal
  acceptance. The signal region (blue, close to the GC) is constructed
  as intersection of the annulus and a circle around the GC with
  radius  $\Delta_{\rm{cut}}$. The remaining regions on the
  annulus (red) are used as background region. Regions within $\pm
  0.3^{\circ}$ of the galactic  plane (yellow) are neither part of the
  signal nor of the background region. }
\end{figure}

\begin{table}[h!t]
\centering
\begin{tabular}{c|cccc}
\hline
Array & $b$ & $r_1$ & $r_2$ & $\Delta_{\mathrm{cut}}$  \\
\hline
E & $1.42^{\circ}$ & $0.55^{\circ}$ & $2.88^{\circ}$ & $1.36^{\circ}$\\ 
B & $1.40^{\circ}$ & $0.44^{\circ}$ & $2.50^{\circ}$ & $1.29^{\circ}$\\ 
\hline
\end{tabular}
\caption{\label{tab:halo_ring_param}Optimized values of the parameters used in the application
of the {\em Ring Method} for the candidate arrays $E$ and $B$. See Fig.~\ref{fig:halo_rings}
for a description of the parameters.}
\end{table}

Given an observation time $T$, Eq.~\ref{eq:halo_rate} can be used to 
estimate the number of excess photons for a particular realization of CTA
and a DM model defining  $\left(\sigma_{\rm ann} v\right)$, $dN_{\gamma}/dE_{\gamma}$ and $J(\psi)$.
Equivalently, one can place a limit on $\left(\sigma_{\rm ann} v\right)$ given 
an upper limit on the number of excess photon events. 
Simulations of the candidate
arrays $E$ and $B$ at a zenith angle of $20^{\circ}$ were used to infer the
effective area for diffuse photons and the residual rate of protons
anywhere in the FOV. Both arrays feature large-size
telescopes and are therefore 
suitable for studies in the low-energy domain.  The available
observation time was set to 100\,h, which is about 10\,\% of the total
observation time per year.

Two different
ways of defining signal and background regions were employed and 
compared, namely the so-called {\em Ring Method} and the On-Off Method.
For the {\em Ring Method}, the candidate arrays
$E$ or $B$ were assumed to observe the GC region at Galactic longitude
$l=0$ and Galactic altitude $b$, and signal and background regions 
were placed in the same FOV as illustrated in Fig.~\ref{fig:halo_rings}. 
An annulus with inner radius $r_1$ and outer radius $r_2$ around the observation
position was constructed and divided into signal and background region
such that the signal region is closer to the GC and has therefore 
a larger astrophysical factor. The separation of signal and background 
region is achieved by a circle with 
radius $\Delta_{\mathrm{cut}}$ around the GC whose intersection with the
annulus defines the signal region. All other regions on the ring were
considered as background region. The values of the four parameters 
$b$, $r_1$, $r_2$ and  $\Delta_{\mathrm{cut}}$ were optimized such that 
the attained significance of a DM signal per square root time was
maximized. The maximization was carried out for a wide range of
WIMP masses but the dependence on the actual WIMP mass was found to be 
fairly weak. The derived values for both candidate arrays are listed
in Tab.~\ref{tab:halo_ring_param}. Judging from present IACT
observations, we do not expect strong diffuse gamma-ray emission to
extend outside the $\pm0.3^\circ$ box used to mask the galactic disc. New
point-like or slightly extended sources will be excluded, making the
On and Off region smaller. In addition, the approach is only sensitive
to gradients in the diffuse gamma-ray emission, whereas the charged
particle background is isotropic. In the optimization process an
Einasto profile was assumed for the DM signal, but the optimal values
are only weakly dependent on the assumed profile in the region beyond
$0.3$ degrees from the Galactic plane. 

The usage of the annulus implies the same acceptance for signal and
background region since the acceptance is, to good approximation, only
a function of the distance to the  observation position. Placing both
signal and background regions in the same FOV implies that
both regions will be affected by time-dependent effects in a similar
way. A disadvantage is, however, that the angular distance between the
signal and background region is only of order of the FOV diameter,
reducing the contrast in Eq.~\ref{eq:halo_rate} considerably. This contrast was increased in
the On-Off Method where data-taking with an offset of typically 30' in Right Ascension
was assumed. In this mode, the telescopes first track
for half an hour the same observation position as in the {\em Ring Method} which 
defines the signal region. The telescopes then slew back and follow
the same path on the sky for another 30\,min. 
The second pointing has the same acceptance as the first one since
the same azimuth and zenith angles are covered but generates a 
background region with much increased angular distance to the GC. 
In the On-Off Method, the observation time was 50\,h for the signal
and 50\,h for the background region giving again a total observation
time of 100\,h. Regardless of whether the {\em Ring Method} or the
On-Off Method was used, all 
areas with $|b|<0.3^{\circ}$ were excluded from signal and background
regions to avoid pollution from astrophysical  gamma-rays.
\begin{table}[h!t]
\centering
\begin{tabular}{l|cccc}
\hline
  &   &  & &  \\[-3.5mm]
Method & Array & $m_\chi$ &  $\widetilde{J}_{\mathrm{s}}$ & $\Delta\Omega$ \\
       &       & (TeV)      &  ($10^{22}\,\mbox{GeV}^2\,\mbox{cm}^{-5}$) & (sr) \\
\hline
Ring    & E & any & 4.68 & 0.00117 \\
        & B & any & 4.43 & 0.00104 \\
\hline
On-Off  & E & 0.1 & 16.4 & 0.00751 \\
        &   & 1   & 19.7 & 0.01044 \\
        &   & 10   & 28.7& 0.02211 \\
\hline
On-Off  & B & 0.1 & 16.4 & 0.00751 \\
        &   & 1   & 22.8 & 0.01384 \\
        &   & 10   & 28.7 &0.02211 \\
\hline
\end{tabular}
\caption{\label{tab:halo_jfactor}Astrophysical factor for the signal 
region and size of the integration region ($\Delta\Omega$) for {\em Ring} and On-Off method. 
In the case of the On-Off Method, $\Delta\Omega$ was chosen as
the entire FOV of the candidate array which introduces
a dependence on the assumed WIMP mass since the effective FOV  grows with photon energy. The table gives values for a WIMP
mass of 0.1, 1, and 10\,TeV.
}
\end{table}

The astrophysical factor (Eq.~\ref{eqn:jbar}) was taken from the Aquarius Simulation~\citep{Aquarius2008}
which had been corrected for the presence of subhalos below the resolution 
limit of the simulation. The line-of-sight integral assumes a value 
of $40.3\times 10^{24}\,\mbox{GeV}^2\,\mbox{cm}^{-5}\,\mbox{sr}^{-1}$ 
at $\Psi=1^{\circ}$. Table~\ref{tab:halo_jfactor}
lists the astrophysical factors of the signal regions
which were defined in the {\em Ring} and On-Off Method, respectively. 
In case of the On-Off Method, the signal region was defined
as the total effective FOV of the On--pointing which
introduces a dependence on the WIMP mass since the FOV
grows with photon energy. For the WIMP annihilation spectrum $dN_{\gamma}/dE_{\gamma}$ 
several different choices were considered. The generic Tasitsiomi spectrum \citep{Tasitsiomi2002} is appropriate for 
a dominant annihilation into quark-antiquark pairs with subsequent 
hadronization into $\pi^0$ particles and was used in the optimization 
of the parameters of the {\em Ring Method}. Other spectra were explored
by considering $b\bar{b}$, $\tau^+\tau^-$ and $\mu^+\mu^-$ final states
\citep{Cembranos:2010xd}. 
%

\subsection*{Discussion}
\begin{figure}[h!t]
\centering
\includegraphics[width=0.95\linewidth]{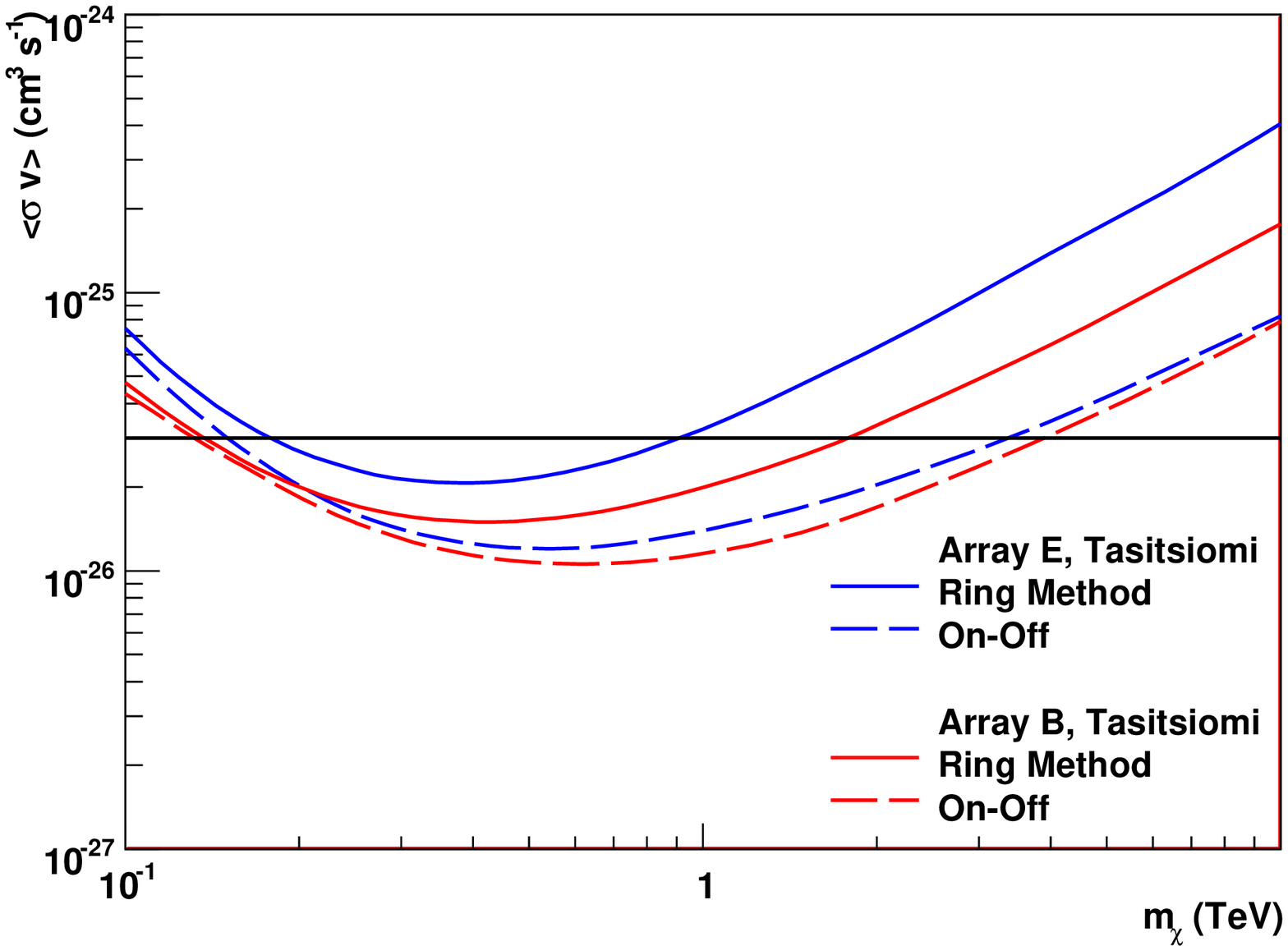}
\includegraphics[width=0.95\linewidth]{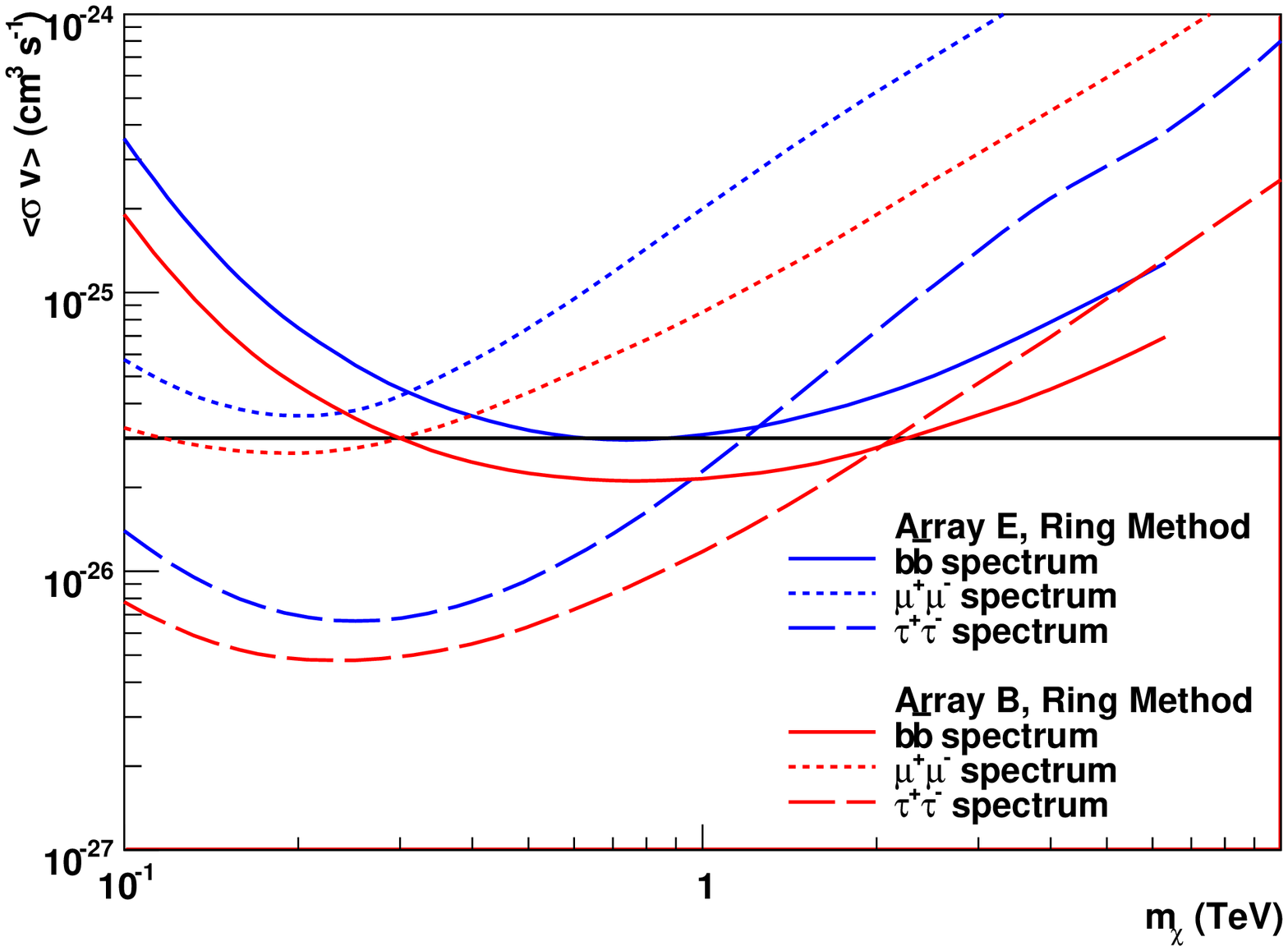}
\caption{\label{fig:halo_methods_spectra}
CTA sensitivities on the velocity averaged annihilation
  cross-section as a function of the WIMP mass. Shown are curves for
  the candidate arrays $E$ (blue) and $B$ (red).  {\bf Top:} Comparison of
  the {\em Ring Method} (solid lines) and On-Off Method for background
  subtraction. Annihilation as in \citet{Tasitsiomi2002} was assumed.
  {\bf Bottom:} Comparison of different WIMP spectra for the {\em Ring Method}. 
  The solid line denotes the case of annihilation into $b\bar{b}$;
  $\mu^+\mu^-$ and $\tau^+\tau^-$ spectra are shown by the
  dotted and dashed lines, respectively. On both panels, the classical
  annihilation cross section for thermally produced WIMPs at
  $3\times10^{-26}\,\mbox{cm}^3\mbox{s}^{-1}$ is indicated
  by the black horizontal line. 
}
\end{figure}

The two plots in Fig.~\ref{fig:halo_methods_spectra} show the upper
limits for WIMP masses between 0.1\,TeV and 10\,TeV, translated from
the sensitivity using here the method
of~\citet{FeldmanCousins1998}.  Each curve corresponds to one set of
assumptions. It is evident that the most constraining limits  
can be derived for masses of about 0.5\,TeV which is
a factor of 2 improvement compared to current IACT arrays like
H.E.S.S. reaching best sensitivity around 1\,TeV. This is a direct 
consequence of the lower threshold and superior stereoscopic background
rejection of the CTA candidate arrays. Typical limits are around 
few $10^{-26}\,\mbox{cm}^3\mbox{s}^{-1}$ which is a factor of 10 improvement
compared to current IACTs. The comparison of array $E$ (blue) and $B$ (same
line style but red) shows that the limits for array $B$ are always 
better, which can be understood from the fact that $B$ comprises $5$
large-size telescopes
and array $E$ only $4$. The magnitude of this effect is, however, comparatively 
small ($\sim 20\,\%$). Overall, CTA should be able to probe
the parameter space below the velocity averaged annihilation cross-section
for thermally produced DM of $3\times 10^{-26}\,\mbox{cm}^3\mbox{s}^{-1}$ for
WIMP masses between several ten GeV and several TeV.

The upper panel of Fig.~\ref{fig:halo_methods_spectra} illustrates the
impact of data-taking with the {\em Ring Method} and the On-Off Method for
the case of a dominant annihilation into quark-antiquark pairs with subsequent
$\pi^0$ creation \citep{Tasitsiomi2002}. 
The On-Off Method (dashed lines) is more sensitive than the {\em Ring
Method} (dashed lines). One must keep in mind, however, that the On-Off Method spends 50\,\% of the observation time
far away from the GC which implies that this data set will be of limited use for studies of
astrophysical sources. Another drawback of the On-Off Method is its susceptibility 
to systematic effects arising from variations of the data-taking conditions
(electronics, atmosphere). In view of this, the increased sensitivity
for the DM halo analysis in parts of the parameter space will not probably suffice to motivate the acquisition 
of a larger data set in this mode.

Compared with the choice of the CTA candidate array ($B$ or $E$) and
the analysis method ({\em Ring Method} or On-Off Methods), the WIMP
annihilation spectrum has the strongest impact on the CTA
sensitivity. The lower panel of Fig.~\ref{fig:halo_methods_spectra}
shows for both candidate arrays and the {\em Ring Method} the limits
obtained in the case of a dominant annihilation into $b\bar{b}$ pairs
(solid), $\mu^+\mu^-$ (dotted) and $\tau^+\tau^-$ (dashed). The small
photon yield from $\mu^+\mu^-$ final states implies limits that are a
factor of about 5 worse than limits for dominant annihilation into
$\tau^+\tau^-$.  It is clear that the full potential of the halo analysis  
will be exploited by confronting individual DM models with their
predicted WIMP annihilation spectra $dN_{\gamma}/dE_{\gamma}$ with data.

%
%
%
\subsection{Anisotropies in the diffuse gamma-ray background\label{sec:spatial}}
Besides gamma-rays from individual resolved sources and Galactic
foreground, another component of diffuse gamma-ray background
radiation has been detected and proven to be nearly isotropic. This
radiation dominantly originates from conventional unresolved point
sources below the detection threshold, while another fraction might be
generated by self-annihilating (or decaying) DM particles, which then
could produce specific signatures in the anisotropy power spectrum of
the diffuse gamma-ray background~\cite{Ando:2007, Ando:2009,
  Cuoco:2008, Siegal-Gaskins:2008, Siegal-Gaskins:2010}. The different
hypotheses about the origin of the gamma-ray background may be
distinguishable by accurately measuring its anisotropy power spectrum.

Compared to the current generation of IACTs, CTA will have improved
capabilities to measure anisotropies in the diffuse gamma-ray
background, based upon a better angular resolution (determined by the
point-spread function, PSF), an increased size of the FOV, and a higher
background rejection efficiency. In the following, we discuss the
effects of different assumptions on the background level and the
anisotropy spectrum on the reconstruction of the power spectrum for
the current generation of IACTs, and address the improvement
obtainable with CTA. Finally, we make predictions for the
discrimination between astrophysical and dark matter induced
anisotropy power spectra for CTA.

\subsection*{Simulation}
In order to investigate the measured power spectrum and the impact of
instrumental characteristics, a sample of \emph{event lists}
containing anisotropies generated with Monte-Carlo simulations was
analyzed. The event lists were simulated by generating skymaps with a
given anisotropy power spectrum. In total, $12$ skymaps covering the
size of the FOV and being in different celestial positions were
created, with a power spectrum for a given multipole moment $\ell$
defined as $C_{\ell} = 1/(2\ell+1) \sum \lvert a_{\ell m} \rvert^2$,
$m = -\ell, \dots, \ell$, where $a_{\ell m}$ denotes the coefficients
of a (real-valued) spherical function decomposed into spherical
harmonics. With $\langle a_{\ell m} \rangle = 0$, $C_\ell$ reflects
the width of the $a_{\ell m}$ distribution, which was assumed to be
Gaussian. The simulations were made for different power spectra $\ell
(\ell +1) C_{\ell} \sim \ell^{s}$, with $s =
0.5,~1.0,~1.5,~2.0,~2.5$. The pixel size of these skymaps was
$0.002^{\circ}$, corresponding to $\ell = 9 \times 10^{4}$ (where
$\Theta_\ell = 180^\circ/\ell$). The skymaps $I(\vartheta, \varphi)$
were normalized in a way that the pixel with the smallest signal was
assigned the value~$0$ and the pixel with the largest signal was
assigned $1$. Anisotropy power spectra were then derived from the
fluctuation maps $I(\vartheta, \varphi)/\langle I \rangle$, such that
for a \textit{full signal} the maximum allowed difference in each map
equals $1$. Note that this difference can be smaller when an
additional isotropic \textit{noise} component is present.

An event was simulated in three subsequent steps: First, the celestial
position was randomly chosen within the FOV, and the event was
classified to represent a \textit{signal}- or isotropic
\textit{noise}-event, respectively. The decision for a signal event
was based upon a normalized random number $z$: If $z$ was smaller than
the skymap value at the corresponding position, the event was
considered a signal event. Otherwise, another event position was
selected while reapplying the procedure. Subsequently, the event map
was convolved with a PSF of $0.1^\circ$, which is similar to the
resolution of current IACTs. The effect of a better angular resolution
is discussed below. The event maps were simulated to contain $10^{7}$
entries. Note here that this number, as selected for the toy model,
does in general not reflect the actual number of expected physical
signal events. Therefore, the following discussion is focussed more on
a qualitative discussion of the criticalities of the calculation
rather than on making quantitative predictions.

To analyze an event list containing $N_{\rm{ev}}$ events, a HEALPix
skymap with $N_{\rm{pix}}$ pixels was accordingly filled, and analyzed
using the HEALPix software
package\footnote{http://healpix.jpl.nasa.gov/}. Therefore, the
analyzed function is
\begin{equation}
  w(\hat{n}) f(\hat{n}) = \frac{N_{\text{pix}}}{N_{\text{event}}} 
  \sum_{i = 1}^{N_{\text{pix}}} x_{i} \cdot b_{i}(\hat{n}),
\end{equation}
where $x_{i}$ denotes the number of events in pixel $i$, and
$b_{i}(\hat{n})$ equals $1$ inside pixel $i$ and $0$ outside. The
function $w(\hat{n})$ describes the windowing function --- in this
case the FOV with Gaussian acceptance --- and $f(\hat{n})$ denotes the
original signal function over the full sky. The windowing function was
normalized such that the integral over the full sky equals $4 \pi$:
\begin{equation} \label{eq_norm}
\int d\Omega \, w(\hat{n}) f(\hat{n}) = 4 \pi.
\end{equation}
%
%
Note that this differs from other analyses of this type, where $w$
is defined such that the maximum value is $1$. This difference in the
normalization was done in order to keep a simple simulation code, and
results should be equivalent. Final results were averaged over the
corresponding $12$ skymaps.

\subsubsection*{The effect of the anisotropy spectrum and the
  residual background level on the spectral reconstruction}

\begin{figure}[h!t]
  \centering
  \includegraphics[width=0.95\linewidth]{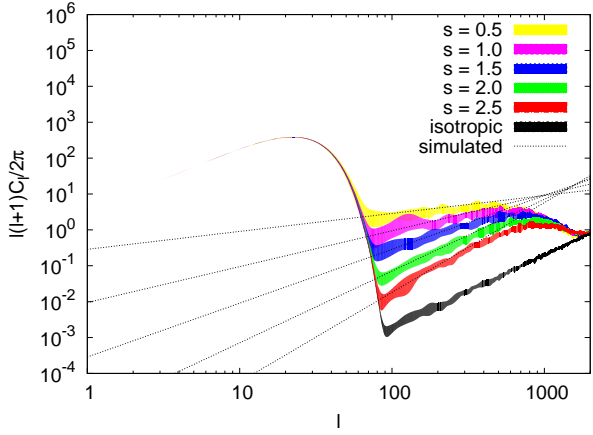}
  \caption{\label{fig_Psf}Measured power spectra $l(l+1)C_\ell/(2\pi)$
    for different slopes $s$ of the simulated input spectrum, compared
    to an isotropic background spectrum. Color-filled areas depict the
    RMS of the spectra. The size of the PSF is $\sigma_{\mathrm{PSF}}
    = 0.1^{\circ}$. For reference, the simulated spectra are shown as
    dashed lines.}
\end{figure}

In Fig.~\ref{fig_Psf}, we show the mean value and the RMS of the
$C_{\ell}$ power spectra. The value $C_{\ell}$ represents the strength
of anisotropies of the angular scale $\Theta_{\ell} =
180^{\circ}/\ell$. Anisotropies smaller than the angular resolution
(defined by the PSF) are smeared out. This effect is clearly visible
for large $\ell\geq1000$, where the power spectra converge into the
Poissonian noise of the isotropic background spectrum. The angular
resolution assumed for the simulation shown in this figure has a width
of $\sigma_{\mathrm{PSF}} = 0.1^{\circ}$. Furthermore, anisotropies
with a size larger than the FOV are truncated at $\ell\sim100$ due to
the effect of the windowing function. The simulated FOV in
Fig.~\ref{fig_Psf} has a width of $2.5^{\circ}$, which is comparable
with the FOV of current IACT experiments. For the toy model,
Fig.~\ref{fig_Psf} demonstrates that, for $\ell \sim 100\!-\!1000$,
power spectra of different slopes are separable within the statistical
errors and distinguishable from isotropic noise. CTA will have a
smaller PSF as well as a larger FOV. This will make the signal vanish
at larger $\ell$ than in the example, and the windowing function will
influence the spectrum to smaller $\ell$ than in the
figure. Therefore, we conclude that the FOV as well as the PSF, while
important, will not be crucial for the investigation of anisotropies
with CTA in the desired multipole range.

\begin{figure}[h!t]
  \centering
  \includegraphics[width=0.95\linewidth]{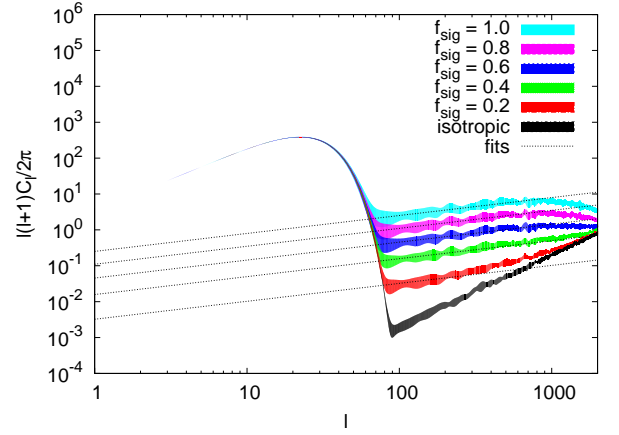}
  \caption{\label{fig_sn}Influence of the signal fraction on the
    measured power spectrum. Shown are the reproduced spectra for a
    slope $s = 0.5$ for several ratios between signal and total
    events; the \textit{background} events are distributed
    isotropically. In order to estimate the effect of the noise ratio,
    the best fit levels are shown as dotted lines. The width of the PSF
    is chosen as in Fig.~\ref{fig_reality}, $\sigma_\mathrm{PSF}
    = 0.05^\circ$.}
\end{figure}

In general, the measured flux will be composed of both signal and
background events. The background is produced mainly by
two separate processes: 
\begin{enumerate}
\item Events caused by cosmic rays (protons and electrons) which are
  misinterpreted as photon events. 
\item An isotropic component of the photon background radiation, which
  does not count as \textit{signal} according to our definition.
\end{enumerate}
The influence of isotropic background is demonstrated in
Fig.~\ref{fig_sn}, where the power spectrum for $s = 0.5$ is shown for
different background levels. Here, the signal fraction is defined by
$f_\mathrm{sig} = N_\mathrm{sig}/N_\mathrm{ev}$, where
$N_\mathrm{sig}$ denotes the number of signal events. The overall
power is clearly reduced in case of fully isotropic background.  From
the figure, we see that when the signal fraction improves by a factor 5, the power spectrum is boosted by about two orders of
magnitude. For this reason, we expect the ten-fold improved CTA
sensitivity to mark the major difference with respect to the current
generation of IACTs for such studies.

\subsubsection*{Prospects for astrophysical and dark matter
  anisotropies discrimination}
\begin{figure}[h!t]
\begin{center}
\includegraphics[width=0.95\linewidth]{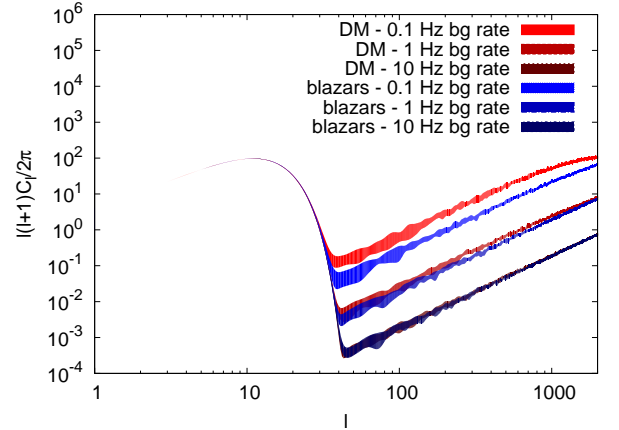}
\end{center}
\caption{Comparison between measured anisotropy power spectra with (a)
  a pure astrophysical origin represented by $C_{\ell} = 10^{-5}$
  (blue bands) and (b) with an additional DM component, i.e., 20\% of
  the total flux, represented by $C_{\ell} = 10^{-3}$ (red bands). The
  assumed observation time is $300\,\mathrm{h}$. The three cases in
  each plot represent the hadronic background rates of
  $10\,\mathrm{Hz}$,$1\,\mathrm{Hz}$, and $0.1\,\mathrm{Hz}$.}
\label{fig_reality}
\end{figure}

The theoretical expectations for the power spectra of the diffuse
gamma-ray flux of both the astrophysical as well as the DM components
are highly model dependent. Since the astrophysical component is
dominated by the gamma-ray flux from unresolved point sources,
expected with a constant $C_{\ell}$ ($s = 2.0$ in our notation), we
conservatively assume the slope of the DM component(s) to be
similar. In this scenario, the difference between the power spectra
manifests in the normalization. For unresolved point sources,
$C_{\ell,\,\mathrm{blazars}} = 10^{-5}$, while for DM-induced
anisotropies, considering the thermal annihilation cross-section
$3\times10^{-26}$~cm$^{3}$ s$^{-1}$, $C_{\ell,\,\mathrm{DM}} =
10^{-3}$ is expected~\cite[see, e.g.,][]{Siegal-Gaskins:2010}. In our
simulation, this was realized by distributing $N=4 \pi/C_{\ell}$ point
sources over the full sky. While representing a non-physical model,
this is a convenient way of producing a Poissonian anisotropy power
spectrum which is a reasonable assumption for generic astrophysical
and DM emitters. The normalization of the signal was set by
extrapolating the spectrum of the extragalactic gamma-ray background
(EGB) \cite{biB_Fermiiso} to $E > 100\,\mathrm{GeV}$. Note that the
strength of the DM annihilation signal is strongly affected by the
formation histories of DM halos and the distribution of DM
subhalos. For example, Fig. 3 in \cite{biB_Fermilimit} shows that the
gamma-ray spectrum of DM annihilation could reach the measured
gamma-ray background spectrum and therefore deliver a significant
fraction of the measured flux. Here, we investigate the cases that
$(a)$ the total EGB originates from astrophysical sources and $(b)$
20\% of the EGB (optimistically) originates from DM annihilation. The
isotropic hadronic component depends on analysis cuts and the quality
of the gamma-hadron separation. In the following, three different
background rates are assumed: $10\,\mathrm{Hz}$, $1\,\mathrm{Hz}$, and
an optimistic $0.1\,\mathrm{Hz}$ rate. We assume a CTA-like FOV with a
radius of $5^\circ$ and a CTA-like PSF with $\sigma_\mathrm{PSF} =
0.05^\circ$. The results are shown in Fig. \ref{fig_reality}, where each
band represents a sample of 20 realizations. One can see in the figure
that depending on the achieved background rate, in principle the two
above mentioned scenarios $(a)$ and $(b)$ will be well distinguishable
for CTA.

%
%
%
\section{Search of axion-like particles with CTA\label{sec:axions}}
Axions were proposed in the 1970's as a by-product of the
Peccei-Quinn solution of the strong-CP problem in QCD~\cite{PQ}. In addition, they are valid candidates to constitute a portion of,
or perhaps the totality of,  the non-baryonic CDM content predicted to exist in the
Universe. Another extremely interesting property of axions, or more
generically, Axion-Like Particles (ALPs, for which --- unlike axions
--- the mass $m_a$ and the coupling constant are not related to each
other), is that they are expected to convert into photons (and
\emph{vice versa}) in the presence of magnetic fields \cite{dicus,
  sikivie}. The photon/ALP mixing is indeed the main signature used at
present in ALP searches, such as those carried out by CAST
\cite{Andriamonje:2007, Arik:2008mq, Aune:2011rx, Iguaz:2010} or ADMX
\cite{admx}, but it could also have important implications for
astronomical observations. For example, photon/ALP mixing could
distort the spectra of gamma-ray sources, such as Active Galactic
Nuclei (AGN) \cite{hooper,deangelis,hochmuth,simet} or galactic
sources, in the TeV range \cite{mirizzi07}.

\begin{figure}[h!t]
\centering
\includegraphics[width=0.95\linewidth]{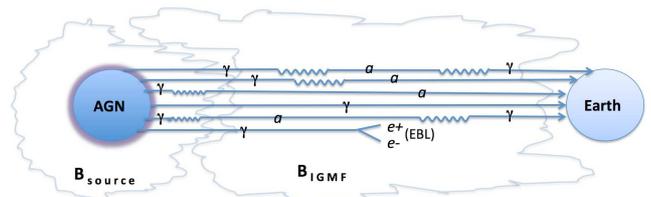}
\caption{\label{fig:sketch}Photon/ALP conversions (crooked lines) that
  can occur in the emission from a cosmological source. {\it $\gamma$}
  and {\it a} symbols represent gamma-ray photons and ALPs
  respectively. This diagram collects the main physical scenarios that
  we might identify inside our formalism. Each of them are
  schematically represented by a line that goes from the source to the
  Earth. From~\citep{alps_Masc}.}
\end{figure}

The photon/ALP mixing effect for distant AGN was also evaluated by
\citet{alps_Masc} under a consistent framework, where mixing takes
place inside or near the gamma-ray emitter as well as in the
intergalactic magnetic field (IGMF). A diagram that outlines this
scenario is shown in Fig.~\ref{fig:sketch}. The artistic sketch shows
the travel of a photon from the source to the Earth and the main
physical cases that one could identify\footnote{Note that this
  formalism neglects, however, the mixing that may happen inside the
  Milky Way due to galactic magnetic fields. In the most
  idealistic/optimistic case, it would produce a photon flux
  enhancement at Earth of~$\sim$3\% \cite{simet}.}. From top to
bottom: $1)$ the photon converts to an axion and back to photon in the
IGMF, $2)$ the photon converts to an axion in the IGMF, $3)$ the photon
converts to an axion at the source, which then does not interact with
the EBL, therefore traveling unimpeded from the source to the Earth, $4)$
the photon travels unimpeded from the source to the Earth, $5)$ the
photon converts to an axion at the source and back to photon in the IGMF,
$6)$ the photon interacts an EBL photon resulting in pair
production. It is clear that 
cases $2,\,3,\,6$ corresponds to an attenuation of the intrinsic
source flux, while cases $1,\,4,\,5$ allow for a recovery of the
intrinsic photon yield.

The probability of a photon of energy $E_{\gamma}$ to be converted
into an ALP (and \emph{vice versa}) can be written as \cite{hooper}: 
\begin{equation}\label{eq:prob2}
P_0 =\frac{1}{1+(E_{crit}/E_{\gamma})^2}~
\sin^2\left[\frac{B~s}{2~M}\sqrt{1+\left(\frac{E_{crit}}{E_{\gamma}}\right)^2}\right]
\end{equation}
\noindent where $s$ is the length of the domain where there is a
roughly constant magnetic field $B$, and $M$ the inverse of the
coupling constant. Here we also defined a characteristic energy,
$E_{crit}$:
\begin{equation} 
E_{crit} \equiv \frac{m^2~M}{2~B}
\label{eq:ecrit1}
\end{equation}
\noindent or in more convenient units:
\begin{equation}
E_{crit} (GeV) \equiv \frac{m^2_{\mu eV}~M_{11}}{0.4~B_G}
\label{eq:ecrit}
\end{equation}
\noindent where the subindices refer to dimensionless quantities:
$m_{\mu \rm eV} \equiv m/ \mu \rm eV$, $M_{11} \equiv M/10^{11}$ GeV
and $B_G \equiv $ B/Gauss; $m$ is the effective ALP mass $m^2 \equiv
|m_a^2-\omega_{pl}^2|$, with $\omega_{pl}=0.37 \times 10^{-4} \mu \rm
eV \sqrt{n_e/\rm cm^{-3}}$ the plasma frequency and $n_e$ the electron
density. The most recent results from the CAST experiment \cite{Iguaz:2010} give
a value of $M_{11} \geq 0.114$ for ALP mass $m_a \leq 0.02$~eV. At
present, the CAST bound is the most general and stringent limit in the
range $10^{-11}$ eV $\ll m_a \ll 10^{-2}$~eV. \\

The main effect produced by photon/ALP mixing in the source is an {\it
  attenuation} in the total expected intensity of the source just
above a critical energy $E_{crit}$ (see Fig.~\ref{fig:sketch}). As for
the mixing in the IGMFs, despite the low magnetic field B, the
photon/ALP conversion can take place due to the large distances
involved. In the model of \citet{alps_Masc}, it is assumed that the
photon beam propagates over N domains of a given length. The modulus
of the IGMF is the same in all of them, whereas its orientation
changes randomly from one domain to the next, which in practice is
also equivalent to a variation in the strength of the component of the
magnetic field relevant to the photon/ALP mixing.

In discussing photon/ALP conversion in IGMFs, it is also necessary to
consider the important role of the Extragalactic Background Light (EBL), its
main effect being an additional attenuation of the photon flux
(especially at energies above about 100 GeV).  Recent gamma-ray
observations already pose substantial 
challenges to the conventional models that explain the observed source
spectra in terms of EBL attenuation~\cite{aleksic11, aliu09,
  Nesphor1998, Stepanyan2002, Krennrich2008}.

Taken together, photon/ALP conversions in the IGMF can lead to an {\it
  attenuation} or an {\it enhancement} of the photon flux at Earth,
depending on distance, magnetic fields and the EBL model considered. A
flux enhancement is possible because ALPs travel unimpeded through the
EBL, and a fraction of them can convert back into photons before
reaching the observer. Note that the strength of the IGMFs is expected
to be many orders of magnitude weaker ($\sim$nG) than that of the
source and its surroundings ($\sim$G). Consequently, as described by
Eq.~(\ref{eq:ecrit}), the energy at which photon/ALP conversion occurs
in this case is many orders of magnitude larger than that at which
conversion can occur in the source and its vicinity.  Assuming a
mid-value of B$\sim$0.1 nG, and $M_{11}=0.114$ (CAST lower limit), the
effect could be observationally detectable by IACTs only if the ALP
mass is of the order of 10$^{-10}$ eV, i.e. we need ultra-light ALPs.

In order to quantitatively study the effect of photon-axion conversion
over the cosmological distances of AGN, we consider the total
photon intensity.  It becomes then useful to define the {\it axion
  boost factor} as the difference between the predicted arriving
photon intensity without including ALPs and that obtained when
including the photon/ALP conversions. Qualitatively speaking, it is
found that the more attenuating the EBL model considered, the more
relevant the effect of photon/ALP conversions in the IGMF (since any
ALP to photon reconversion might substantially enhance the intensity
arriving at Earth). Furthermore, higher B values do not necessarily
translate into higher photon flux enhancements. There is always a B
value that maximizes the axion boost factors; this value is sensitive
to the source distance, the considered energy and the adopted EBL
model (see Ref.~\cite{alps_Masc} for a more detailed discussion).

There could be indeed different approaches from the observational
point of view, although all of them will be probably based on the
search and analysis of a systematic residual after applying the
best-fit (conventional) model to the AGN data. For example,
Ref.~\cite{alps_Masc} predicts the existence of a universal feature
in the spectrum of the sources due to the intergalactic mixing, that
is completely independent on the sources themselves and only depends
on the ALP and IGMF properties. This feature should be present at the
same critical energy $E_{crit}$ for all sources, and would show up in
the spectra as a drop in the flux --- whenever $E_{crit}$ is in the
range where the EBL effect is negligible --- or even as a sudden flux
increase, if the EBL absorption is strong for $E = E_{crit}$.

\subsection*{Test case for CTA: PKS 1222+21}
We have taken as a test source the flat spectrum radio quasar 4C +21.35
(PKS 1222+21), at redshift $z = 0.432$, which was detected by MAGIC
above 70 GeV \cite{pks1222_MAGIC} in June 2010, during a \emph{target
  of opportunity} observation triggered by the high state of the
source in the Fermi-LAT energy band. This source is the second most
distant object detected by ground-based gamma-ray telescopes, and
hence an ideal candidate for the study of propagation effects. The
observed energy spectrum of 4C +21.35 during the 0.5 hour flare
recorded by MAGIC was well described by a power law of index $\Gamma =
3.75 \pm 0.27_{stat} \pm 0.2_{syst}$. The intrinsic spectrum, assuming
the EBL model of \citet{EBL_Dominguez} was estimated to be a power law
of index $\Gamma = 2.72 \pm 0.34$, which extrapolated down to an
energy of about $5$ GeV, connects smoothly with the harder spectrum
($\Gamma = 1.95 \pm 0.21$) measured by Fermi-LAT between $0.2$ and
$2$~GeV in a $2.5$~h period encompassing the MAGIC observation. It
must be noted that longer-term Fermi-LAT observations of the source in
various states of activity show a break in the spectrum between 1 and
3 GeV, with a spectral index after the break (and up to $\simeq 50$
GeV) ranging between 2.4 and 2.8~\cite{pks1222_Fermi}.

We have simulated CTA observations of 4C +21.35 assuming an intrinsic
unbroken power-law spectrum, in the relevant energy range, like the
one determined by MAGIC for 4C +21.35 during the flare, i.e. $dN/dE =
$ K $\times [E / (0.2$ TeV$)]^{-2.72}$.  Keeping the spectral shape
unchanged, we have tried different absolute flux normalizations,
taking as a reference the flux observed by MAGIC, K $= 1.78 \times
10^{-5}$ m$^{-2}$ s$^{-1}$ TeV$^{-1}$. We have also tested different
observation times: the actual duration of the VHE flare observed by
MAGIC is unknown, since the observation was interrupted while the
flare was still going on, but the flares observed by Fermi-LAT above
100~MeV show rise and decay time scales of the order of a
day~\cite{pks1222_Fermi}, so it is reasonable to expect that the
source may stay several hours in flux states as high as that observed
by MAGIC. For the detector simulation we have used the CTA candidate
array $E$.  The EBL model in Ref.~\cite{EBL_Dominguez} has been used to
account for the effect of the EBL, and the conversion of photons into
ALPs and \emph{vice versa} has been simulated following the formalism
detailed in Ref.~\cite{alps_Masc} as outlined above. Only conversions
in the IGMF have been considered (in this case, mixing in the source
typically leads to only a few percent of flux attenuation, so we
neglected it in order to avoid extra uncertainties).

\begin{figure}[h!t]
\centering
\includegraphics[width=0.95\linewidth]{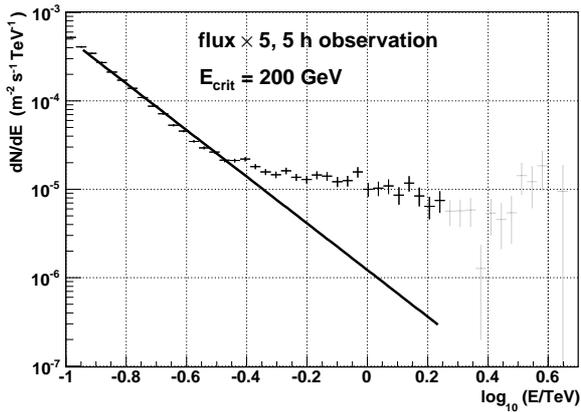}
\caption{\label{fig:Ecrit200}Simulation of a 5~h CTA observation of a
  4C +21.35 flare $5$ times more intense than the one recorded by
  MAGIC~\cite{pks1222_MAGIC}. In black, energy bins used for the fit
  (those with a signal exceeding three times the RMS of the
  background, and a minimum of 10 excess events). Excluded points are
  displayed in grey. The estimated intrinsic differential energy
  spectrum (after correcting for the EBL effect) shows a {\it boost}
  at high energies due to photon/ALP mixing. The IGMF strength is
  assumed to be 0.1~nG, and ALP parameters result in E$_{crit} = 200$
  GeV. }
\end{figure} 

\begin{figure}[h!t]
\centering
\includegraphics[width=0.95\linewidth]{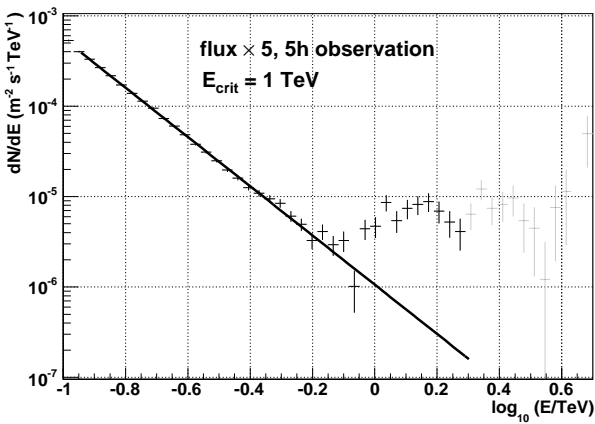}
\caption{\label{fig:Ecrit1000}Same as in Fig.~\ref{fig:Ecrit200}, but
  with E$_{crit} = 1$~TeV. Note that in scenarios like this, where
  E$_{crit}$ is within the energy range in which the EBL absorption is
  already large, the {\it boost} in the flux shows up as a sudden rise
  (smeared out by the spectral resolution of the instrument) which
  would even allow to determine E$_{crit}$ accurately.}
\end{figure} 

We have assumed the same parameters for the IGMF as those in the
fiducial model in~\cite{alps_Masc}: 0.1 nG is the (constant) modulus
of the IGMF\footnote{Note that this is just one order of magnitude
  below the current upper limits \cite{deangelisBfield}, and lower
  limits lie many orders of magnitude below
  \cite{neronov,tavecchio}. With a IGMF of 0.01 nG or smaller the
  effect of the photon/ALP mixing would probably be too weak to be
  observed with CTA.}, which is assumed to have fixed orientation
within domains of size 1 Mpc. The orientation of the IGMF varies
randomly from one domain to the next. The ALP parameters, mass and
coupling constant, enter via $E_{crit}$, below which the conversion
probability is negligible. We have scanned $E_{crit}$ in the range
0.1 to 10~TeV in steps of $0.1$~TeV.

Using the performance parameters of array $E$, we obtain the expected
gamma-ray and cosmic-ray background rates in bins of estimated energy,
and from them the reconstructed differential energy spectrum. After
this, we correct the observed spectrum by the energy-dependent
attenuation factors expected from the EBL in order to get an estimate
of the intrinsic source spectrum. Each simulated spectrum is fitted to
a power-law with variable index of the form $dN/dE \propto
E^{-\alpha-\beta~log(E/0.1\text{TeV})}$, in which we constrain the
$\beta$ parameter so that the spectrum cannot become harder with
increasing energy (such behavior is not expected from emission models
in this energy range). Only energy bins with a signal exceeding three
times the RMS of the background, and a minimum of 10 excess events,
are considered in the fit.

In the absence of any significant photon/ALP mixing, the resulting fits
will all match the spectral points within the experimental
uncertainties, resulting in {\it good} $~\chi^{2}$ values. But, as
shown in Ref.~\cite{alps_Masc}, certain combinations of ALP parameters
and values of the IGMF may result in significant modifications of the
observed VHE spectra. The most striking feature is a {\it boost} of
the expected flux at high energies, which is particularly prominent in
the estimated intrinsic (i.e. EBL-de-absorbed) spectrum. Such a
feature may result in a low value of the $\chi^{2}$-probability of the
spectral fit. In Figs.~\ref{fig:Ecrit200} and \ref{fig:Ecrit1000} we
show two such cases, in which the observed spectra, after
de-absorption of the EBL effect, show a clear hardening of the
spectral index. The effect is particularly striking in the cases in
which the EBL absorption at E = E$_{crit}$ is already strong
(e.g. Fig.~\ref{fig:Ecrit1000}), because then the boost sets in very
fast, resulting in dN/dE rising with energy at around E$_{crit}$. The
rise is actually very sharp, but it is smoothed by the energy
resolution of the instrument. An improvement in the energy resolution
would increase the significance of the feature and improve the
determination of E$_{crit}$. In contrast, if E$_{crit}$ is in the
range in which the EBL absorption is small or negligible
(Fig.~\ref{fig:Ecrit200}), the feature at E$_{crit}$ would just be a
flux drop of at most $\simeq 30\%$ \cite{alps_Masc}, also washed out
by the instrumental energy resolution. In those cases, though a
high-energy boost may still be clearly detected, it would be
hard to determine the exact value of E$_{crit}$. This is because, in the
formalism described in Ref.~\cite{alps_Masc}, similar ALP boost
factors are always achieved at energies $E>E_{crit}$, independently of
the particular value of E$_{crit}$ in each case.

\begin{figure}[h!t]
\includegraphics[width=0.95\linewidth]{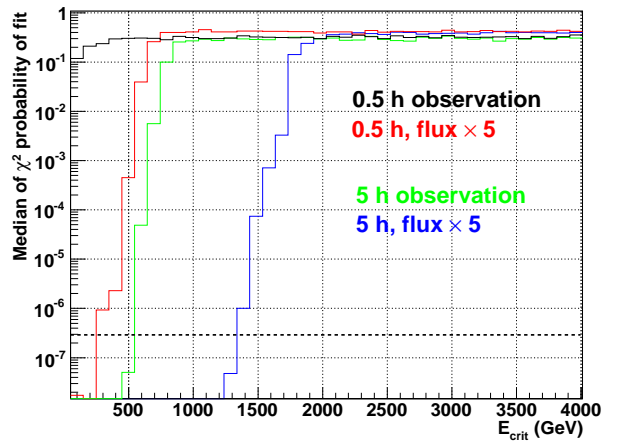}
\caption{\label{fig:probabilities}Median of the $\chi^2$-probabilities
  of the fits to the de-absorbed differential energy spectra of 4C
  +21.35 measured by CTA, assuming photon/ALP mixing, for different
  values of E$_{crit}$.  We simulated observations of flares of two
  different durations: 0.5 and 5 hours, and with intensities equal to
  1 and 5 times that of the flare reported in
  \cite{pks1222_MAGIC}. The dashed horizontal line marks the
  probability that corresponds to 5 standard deviations.}
\end{figure} 

\subsection*{Prospects}
For each of the $E_{crit}$ values scanned, we have performed $10^{3}$
simulations of a CTA observation, all with the same source flux and
observation time. We consider that a given value of $E_{crit}$ is
within the reach of CTA whenever the median of the
$\chi^{2}$-probability distribution is below $2.9\times 10^{-7}$,
which corresponds to 5 standard deviations. In Fig.
\ref{fig:probabilities} we show the median of the $\chi^{2}$
probability versus $E_{crit}$, for two different assumptions on the
source flux and two different observation times. The range of
$E_{crit}$ which can be probed with CTA for the different scenarios is
the one for which the curves in Fig.~\ref{fig:probabilities} are below
the dashed horizontal line. As expected, the range becomes larger as
we increase the observation time and/or the flux of the source. A
0.5~h~duration flare like the one reported in \cite{pks1222_MAGIC}
would not be enough for CTA to detect a significant effect in any of
the tested ALP scenarios, i.e. the solid black line never goes below
the dashed line for any value of E$_{crit}$. A flare of similar
intensity, but lasting 5 hours (green line) would already be enough to
see the boost due to ALPs for those scenarios with E$_{crit} \leq 500$
GeV. In Fig.~\ref{fig:probabilities} we can also see that for a
hypothetical flare with an intensity 5 times larger, lasting 5 hours,
the accessible range of E$_{crit}$ would extend up to 1.3 TeV.

%
%
%

\section{\label{sec:liv}High Energy Violation of Lorentz Invariance}

Lorentz Invariance (LI) lies at the heart of all of modern physics, in
particular the unification of space and time through the principle of
Special Relativity. Space-time was elegantly promoted to be a dynamic
entity in the covariant classical theory of gravity namely General
Relativity (GR) which has been rigorously tested on astronomical
scales and underlies the mathematical description of
cosmology. Similarly quantum mechanics has been successfully married
with Special Relativity to yield the quantum theory of fields which
underlies the very successful Standard Model of leptons and quarks and
the gauged electromagnetic, weak and strong forces. However it has
proved considerably more difficult to unify gravity with the other
forces, since GR is fundamentally non-renormalisable. A fully quantum
theory of gravity (QG) is still beyond our grasp although there has
been significant progress towards this goal in various approaches such
as superstring theory and loop quantum gravity
\cite{Rovelli:2000aw}. QG should describe dynamics at the Planck
energy $E_{\rm Pl}=\sqrt{\hbar c/G_{\rm N}}\simeq 1.22\times
10^{19}$~GeV or equivalently the Planck length $l_{\rm Pl}=\sqrt{\hbar
  G_{\rm N}/c^3} \simeq 1.62\times 10^{-33}$~cm, where gravitational
effects should become as strong as the other forces and the notion of
space-time is likely to need revision. This has opened up the
possibility that LI may be violated by QG effects although, lacking a
fully dynamical theory, the expectation is generic rather than
definite. For example quantum fluctuations may produce `space-time
foam' at the Planck scale resulting in a non-trivial refractive index
and anomalous dispersion of light \emph{in vacuo} i.e. an energy
dependence of the speed of light. Hence over the past decade, there
has been tremendous interest in testing LI at high energies as part of
what has come to be called `quantum gravity phenomenology'
\cite{Sarkar:2002mg,mattingly05,liberati09}.

Possible energy dependence of the speed of light in the vacuum has
been predicted, in the framework of several theories dealing with
quantum gravity models and effective field theory
models~\cite{myers03}. The seminal paper by \citet{Amelino-Camelia98}
proposed that this can be parameterised by a Taylor expansion of the
usual dispersion relation:
\begin{equation}\label{eq:liv}
 c^{2}p^{2} = E^{2}\left[1 \pm \xi_{1}(E/E_{\rm Pl}) \pm \xi_{2}(E/E_{\rm Pl})^{2} \pm
 \ldots \right], 
\end{equation}
where the value of the co-efficients $\xi_\alpha$ would be specified
by the theory of quantum gravity (and may well turn out to be zero) .
For example there are specific predictions in some toy models
\cite{alfaro02,ellis08} and a general parameterisation can be provided
in the framework of effective field theory~\cite{myers03}. For more
details see the introduction by \citet{Ellis:2011ek} in Part~A of this
Special Issue. Typically, two scenarios are envisaged according to
whether the linear term or the quadratic term is dominant,
parametrised by the scale parameters $\xi_{1}$ (linear case) and
$\xi_{2}$ (quadratic case) respectively. The point is that while QG
effects would be prominent only at the Planck scales, there would be
residual Lorentz Invariance Violation (LIV) effects at lower energies (GeV--TeV) in the form of
anomalous photon velocity dispersion.

\citet{Amelino-Camelia98} also noted that over a cosmological distance
$L$, the magnitude of time--delay $\Delta t$ induced by LIV between
two photons with an energy difference $\Delta E$ is detectable:
\begin{equation}\label{eq:LIV_distanceformula}
\Delta t \simeq \left(\frac{\Delta E}{\xi_{\alpha} E_{\rm Pl}} \right)^{\alpha}\frac{L}{c}
\end{equation}
where $\alpha=1$ or $2$ according to whether the linear or quadratic terms
dominates in Eq.~(\ref{eq:liv}). The energy scale of QG is commonly
expected to lie somewhere within a factor $\xi_{\alpha}$ of $E_{P}$.
The best limit on the linear term has recently been placed by
Fermi-LAT observations of GeV photons from GRB 090510 ($z=0.903$)
which require $M_1=E_{\rm Pl}/\xi_1>1.5\times10^{19}$~GeV
\cite{FermiLimits}. The most constraining limit on the quadratic term
$M_2=E_{\rm Pl}/\xi_2>6.4\times10^{10}$~GeV come from observations of
an exceptional flare of the active galactic nucleus (AGN)
PKS\,2155-304 with the H.E.S.S. telescope~\cite{Abramowski:2011jw}.

It is important to keep in mind that although the QG induced
time--delay is proportional to energy (as opposed to conventional
dispersion effects which vary as inverse power of energy) similar
time--delay effects may be intrinsic to the
source~\cite{mastich08}. Therefore, in order to distinguish between
source and propagation time--delays, different types of sources should
be considered with different physical properties and situated at
different cosmological distances. For such studies, AGN and Gamma-Ray
Bursts (GRBs) are
the best candidates to test Eq.~(\ref{eq:LIV_distanceformula}). AGN
cover the higher energies (up to few TeV) and lower redshift regime
(probably up to $z\sim0.8-1$) and GRBs the lower energies (probably
few tens of GeV) but higher redshifts. Other promising candidates
could be pulsars which until now have yielded constraints one order of
magnitude weaker than the ones derived from AGN~\cite{otte11}.


\subsection*{The consequences of improved sensitivity and larger
  energy coverage of CTA on time--delays recovery}

Using the Maximum Likelihood Estimation method (MLE)
of \citet{martinez09}, we investigate the effects that the improved
CTA performance, in terms of increased statistics and broader energy
lever arm, have on the time--delay recovery.

\begin{figure}[h!t]
\centering
\includegraphics[width=0.95\linewidth]{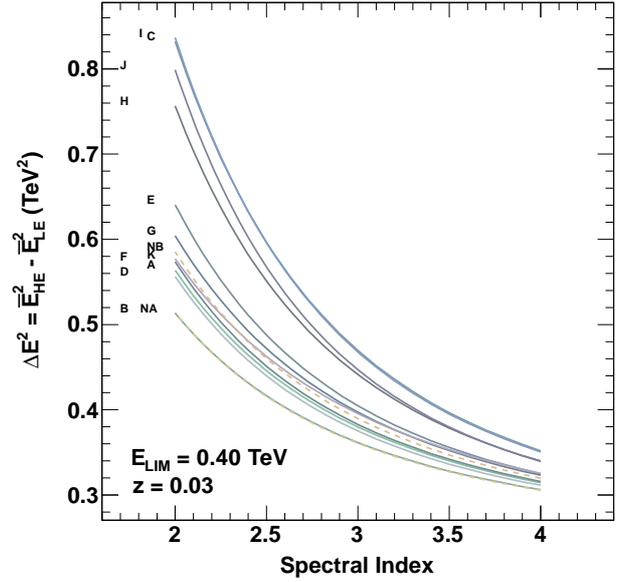}
\caption{\label{fig:jb1}Parameter $\Delta E^2$ as a function of the
  spectral index for all arrays considered in CTA Monte Carlo
  simulations. The configurations have different layouts and number of
  telescopes. See details in the text.}
\end{figure}

Five hundred realisations of Gaussian--shaped ``pulsed'' light curves
were generated for several values of time--delays between
$-60$~s$\,$TeV$^{-1}$ and $+60$~s$\,$TeV$^{-1}$ in steps of 10
s$\,$TeV$^{-1}$. This allowed an estimate of the value of the error
$\delta t_{\rm r}$ on the measured time--delay $\Delta t_{\rm r}$. The
error decreases as $N_\gamma^{-1/2}$, where $N_\gamma$ is the number
of photons included in the likelihood fit, and saturates at a value of
about 3~s\,TeV$^{-1}$, about a factor of 3 less than the current
generation of IACTs, due to the increased statistics of CTA.
The effect of the increase in the energy lever-arm, provided by the
wide coverage of CTA from few tens of GeV to several tens of TeV, has
also been addressed for the different array configurations taking into
account the absorption of Extragalactic Background Light (EBL) using
the model of \citet{kneiske02}\footnote{This is a conservative choice
  --- by using more transparent EBL models like that of
  \citet{EBL_Dominguez}, the results would be more promising.}, and a
spectral break at around 100 GeV. The photons are thus separated into
two populations at ``high'' and ``low'' energies with average values
marked as $\bar{E}_{HE}$ and $\bar{E}_{LE}$ respectively.
Fig.~\ref{fig:jb1} shows the variation of the energy lever-arm for
different CTA configurations in energy-squared (quadratic case)
$\Delta E^2 = \bar{E}_{HE}^2 - \bar{E}_{LE}^2$, after a convolution
with the effective area and for a given choice of the energy value
separating high and low energy bands
$\mathrm{E}_\mathrm{LIM}=400$~GeV. This value is almost stable
regardless of the array or the spectral index~\cite{bolmont11}. The
ranking shows that arrays $I$, $C$, $J$ and $H$ for the southern site and $NB$
for the northern site are favourable for LIV studies.

\begin{figure}[h!t]
\centering
\includegraphics[width=0.95\linewidth]{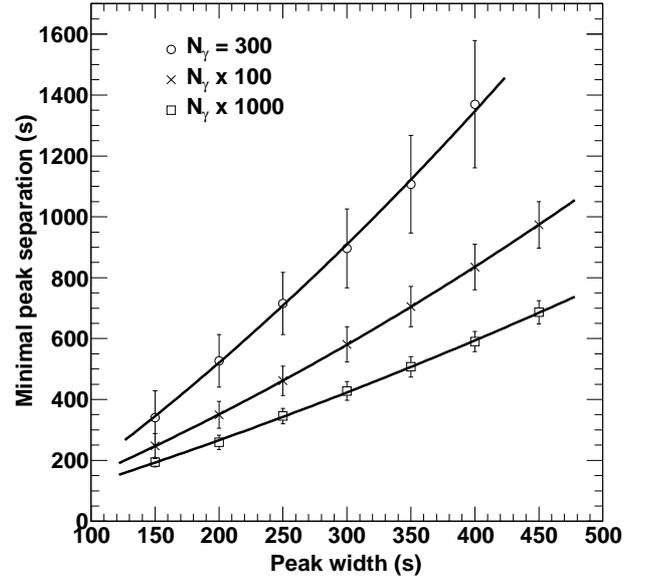} 
\caption{\label{fig:jb2} Minimal distinguishable separation of two
  Gaussian peaks as a function of their width, in case of a
  H.E.S.S.-like measurement (open circles), a CTA-like measurement (crosses) and
  an ideal situtation where a ten times more photons than in the CTA
  measurement could be collected.}
\end{figure}

The intrinsic variability of the photon emission by astrophysical
sources such as GRBs and AGN is the main systematic uncertainty in LIV
searches. Until now, the detected variability of the AGN was limited
to about 100 s, partially due to the limited statistics of the
data. The possibility of improved separation of the initially
unresolved double peak structures was investigated with light curve
simulations and time--delay reconstruction using again the MLE
method. Fig.~\ref{fig:jb2} shows the minimal peak separation which
would allow distinguishing between two Gaussian spikes of the same
standard deviation $\sigma_G$ for different photon statistics: a
H.E.S.S.-like measurement, a CTA-like measurement with an improved
photon collection by a factor 100, and a more optimistic scenario with
a factor 1000 more photons.


\subsection*{Sensitivity to TeV photons for selected benchmark AGN}
Following the suggestion of \citet{Amelino-Camelia98} we define a
sensitivity factor $\eta = |\Delta t_{\rm d}| / T$, where $\Delta
t_{\rm d}$ is the magnitude of the time--delay introduced into the
flare and $T$ is duration of the burst/flare feature that is being
examined. In Refs.~\cite{BarresDeAlmeida:2009,barresdeAlmeida:2010} it
is shown that $\eta \geq 0.3$ is required to determine whether the
observed flare time sequence has been skewed in comparison to its
original form.
To improve upon current limits will require observations of
photons at energies larger than 10~TeV (for a given redshift) or
observations of similar flares from much more distant AGN. In the
following, we calculate the integral numbers of photons from
representative AGN to test LIV signatures in AGN flares.

\begin{figure}[h!t]
  \centering
  \includegraphics[width=0.95\linewidth]{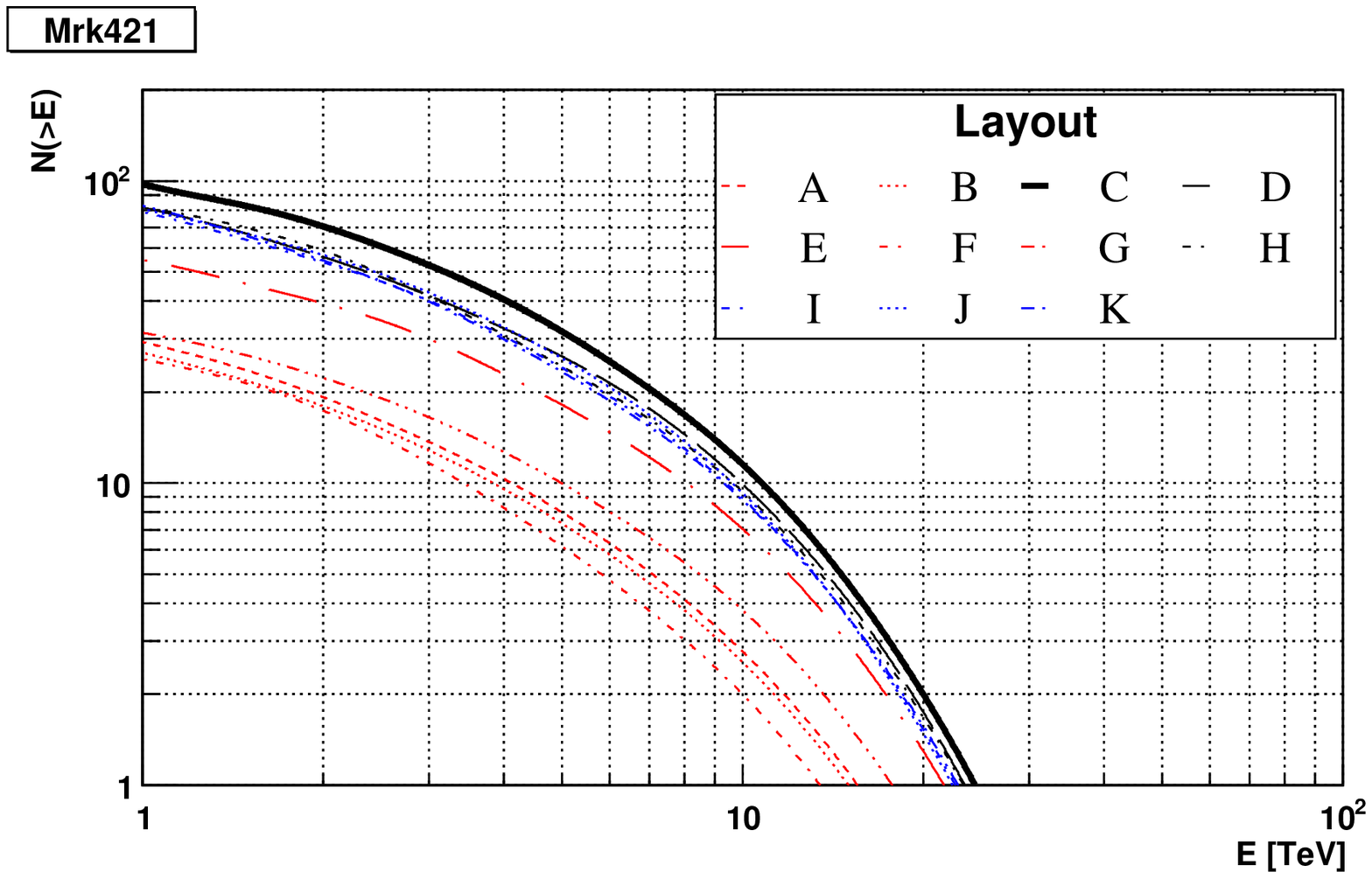}
  \includegraphics[width=0.95\linewidth]{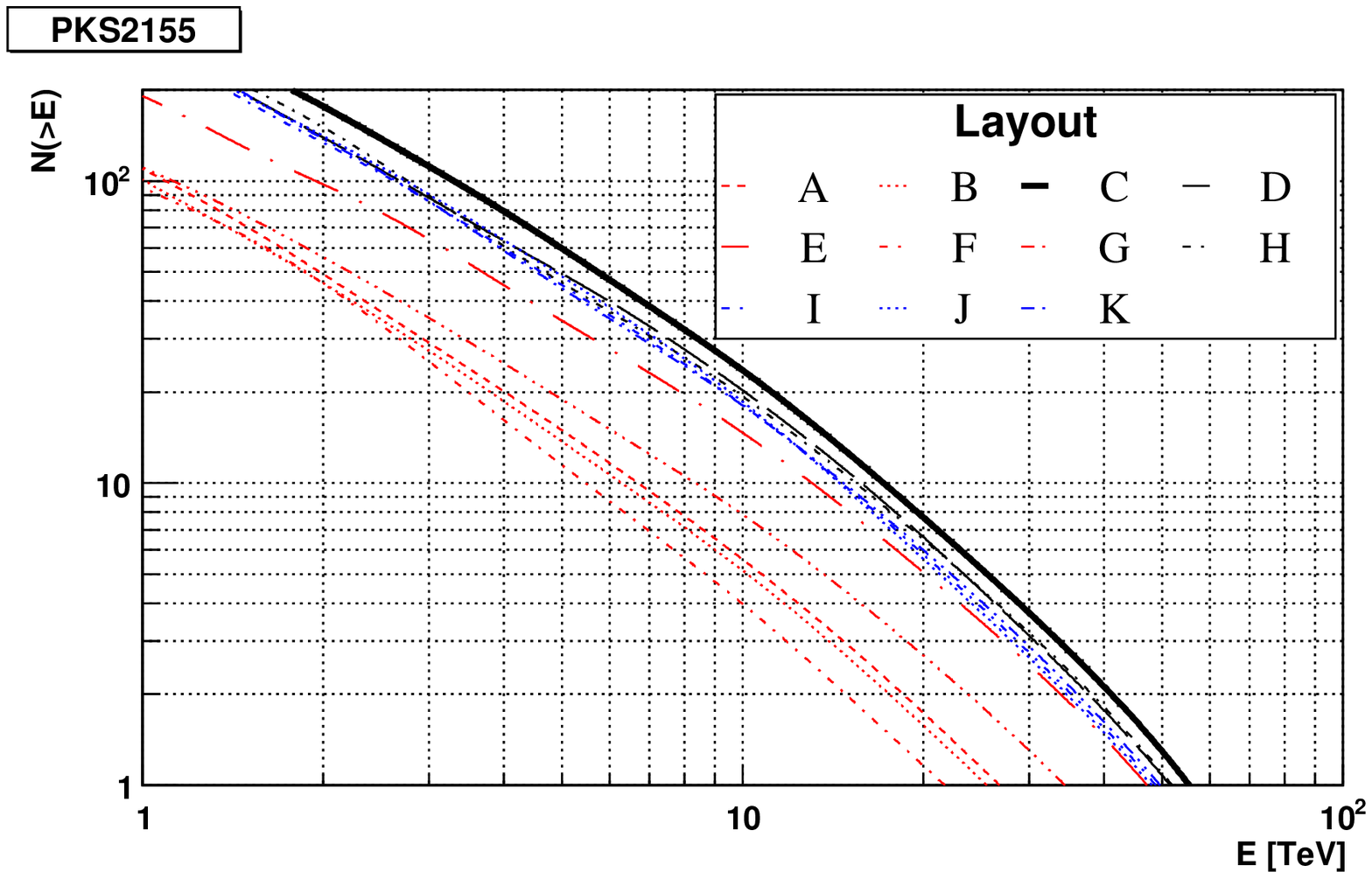}
  \includegraphics[width=0.95\linewidth]{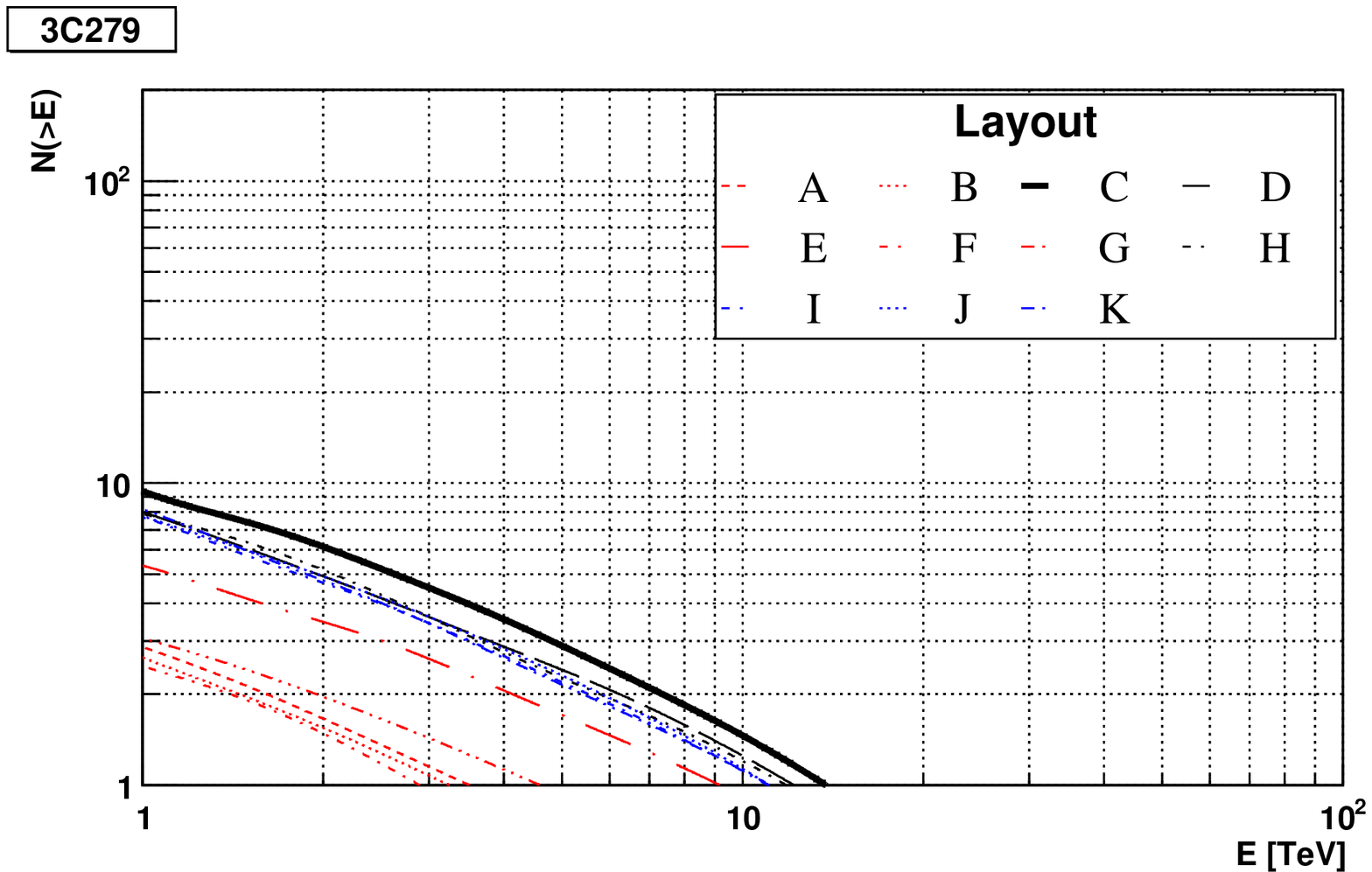}
  \caption{\label{fig:NumberPhotons} Integral number of events above a
    given energy, expected for the various array configurations for
    simulated flares, for $N(>10\,\mathrm{TeV})=10$ photons. Each
    panel is accumulated for the appropriate flare timescale required
    to be able to determine if Planck scale quantum gravity induced
    LIV is present. The top panel is for a 10 Crab Mrk\,421 flare
    \cite{Mrk421} if it lasted for 30\,s duration; the middle panel is
    PKS\,2155-304 similarly at its high level \cite{Aharonian:2007},
    but 120\,s duration; the bottom panel is for a 610s 3C\,279 flare
    at its highest recorded flux level \cite{3C279}.}
\end{figure}

We have taken three VHE AGN representative of several known
situations: an AGN-flare with high brightness (Mrk~421), an AGN-flare
that shows short variability timescales (PKS 2155-304) and the AGN
with the largest known redshift (3C 279) observed at VHE. The spectra
in their highest recorded flux state have been taken from current IACT
observations, extrapolated to higher energies, convolved with the
performance curves of the various CTA array layouts and integrated
assuming a flare duration lasting for the appropriate time such
  that $\eta=0.3$. The results are shown in
Fig.~\ref{fig:NumberPhotons}. Since this falls into the category of
unbinned methodologies, the precision in the time resolution is the
same as the time precision of each array (i.e.\ better than
$1\,\mu$s). The uncertainty in the time--delay for a single photon is
the time precision modulo of the energy resolution (being less than
10\% as specified) and the distance (negligible), and saturates at 10
s\,TeV$^{-1}$\,Gpc$^{-1}$. This translates into a Planck scale effect
of ${\cal O}(10)$\,s (i.e.\ as good as a binned method would have for
more than 100 photons).

Mrk\,421 is known to show spectral hardening with increasing flux, and
can have very hard spectra indeed on short timescales, as evidenced by
Flare C in Ref.~\cite{Mrk421} --- see discussion by
\citet{Gaidos:1996ss}. For the redshift $z=0.03$ of Mrk\,421, Planck
scale effects could be expected to induce a delay of $\sim
1$\,s\,TeV$^{-1}$; for $E\geq10$\,TeV photons this means we would need
to be able to time resolve flare features at an unprecedented
$\sim30$\,s duration. Whilst features this fast have yet to be
identified, this could be because they are below the sensitivity of
current instruments and the top panel of Fig.~\ref{fig:NumberPhotons}
demonstrates that, if present, such features can indeed be probed.
For PKS 2155-304, the redshift $z = 0.117$ implies that CTA would need
features on the timescale of 120~s to test Planck scale effects, which
is still a factor of a few faster than the $240-610$~s rising and
falling timescales of the $\sim$7 Crab flares observed to
date~\cite{Aharonian:2007} but, as shown in the middle panel of
Fig.~\ref{fig:NumberPhotons}, we would easily have sufficient photons
to resolve such features.
For 3C~279 ($z=0.536$,~\cite{3C279}), even though the flare timescale
of 610~s required for such a distant AGN are well within the
variability timescales we currently observe for blazars, the photon
flux we expect ($\sim 10\%$ Crab) from such a distant source is
expected to be too low to resolve such features at the highest
energies, because of the attenuation of the photon flux through
interactions with the EBL.

Concerning the different CTA arrays, Fig.~\ref{fig:NumberPhotons}
shows that the best performing arrays at high energies are $C$, $D$, $H$, $I$,
$J$, $K$. While it may be of more interest to find out if they would
detect sufficient photons on which to perform tests for time--delay, we
note that there are a number of unbinned methods that can cope with
sparse datasets \cite[see, e.g.,][]{BarresDeAlmeida:2009,
  Abramowski:2011jw, DisCan} so that $10$ photons of E$>=$10~TeV are
required in order to be able to begin to test for LIV.

\subsection*{Time--delay recovery with realistic source
  lightcurve and spectra for the linear case} 
\begin{figure}[h!t]
  \centering
  \includegraphics[width=0.95\linewidth]{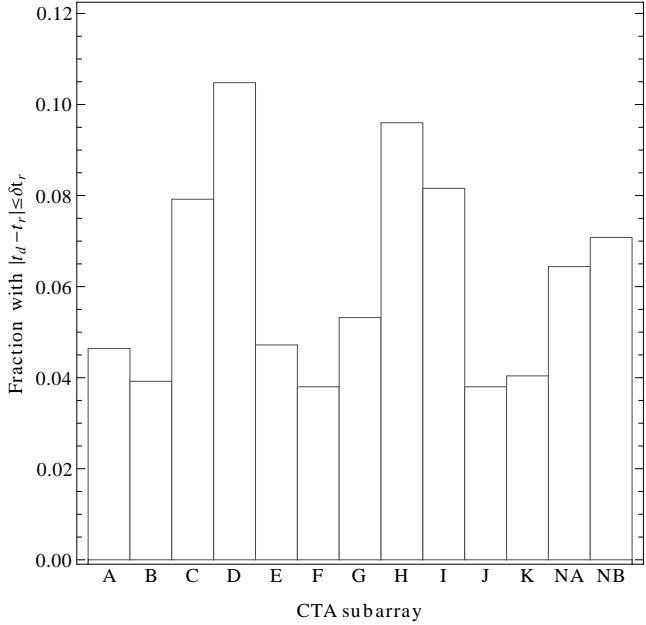}
  \caption{\label{fig:globalAGNHist} Time--delay recovery fraction
    with quality factor $q<1$ for each CTA configuration. See text for
    details.}
\end{figure}

For this study, we produce 2500 pairs of lightcurves (with and
without time--delays), following the method of \citet{timmer95} for
each of the total 13 CTA arrays (11 Southern array $A$\ldots $K$ and 2
Northern arrays $NA$, $NB$), as shown in Fig.~\ref{fig:globalAGNHist}. We
scan the space of possibilities by selecting random values in the
5-dimensional space characterised by the following parameters:
\begin{list}{--}{\itemsep=0pt}
\item time--delays in the range $\xi_{1}=0.1-5$ (linear case)
\item AGN redshift linearly in the range $0.03-0.6$
\item energy spectrum power law slope between 20 GeV
  and 20~TeV in the range $1-2.5$, and with a spectral cutoff at 120~GeV
\item Flux level in the range $10^{-11}-10^{-12}$ ph cm$^{-2}$
  s$^{-1}$
\item different observational periods: a) single day observations
  consisting of 3 pointings of 30 and 15 min, b) weekly observations
  consisting of 3 nightly  pointings of 30 and 15 min, and c) monthly
  observations consisting of   2 nightly pointings of 30 and 15 min.      
\end{list}
The photons thus generated are then distributed as a function of time,
based on the variability type that we have initially assumed
(e.g.\ red-noise).  These light curve pairs incorporate
delay effects accumulated over a given distance depending
on the $\Delta E$ of each pair. The light curve pairs are subsequently
convolved with the CTA arrays performance, using the effective area,
the background count rate and differential sensitivity. Finally, for
each CTA array, we recover the observed time--delays using the
cross-power spectral analysis method of \citet{nowak99}. For each
pair of light curves and for each array we consider the \emph{quality factor}
between the simulated time--delay $\Delta t_{\rm d}$ and the recovered
time--delay $\Delta t_{\rm r}$ defined as:
\begin{equation}\label{eq:qfactor}
q=|\Delta t_{\rm d}-\Delta t_{\rm r}|/\delta
t_{\rm r}, 
\end{equation}
where $\delta t_{\rm r}$ is the width of the Gaussian distribution of
the recovered time--delays coming from the simulations.

When considering all the 13 arrays together, we find on average that
6\% of the time--delays are recovered with $q<1$. This is a very
strict limit. If one relaxes this limit to $q<2$ and $q<3$,
respectively 77\% and 99\% of the time--delays in our sample of events
are recovered, thus making the prospects of detecting (or
constraining) LIV signatures with CTA rather optimistic.

To understand which arrays have the best prospects,
the time--delay recovery results for each array individually is shown
in Fig.~\ref{fig:globalAGNHist}, adopting the limit of
$q<1$.  From the plot we see that the best CTA configurations for
time--delay recovery are $C$, $D$, $H$, $I$ and $NB$ respectively for the
Southern and Northern hemispheres. 

%
%

\section{Other physics searches with CTA\label{sec:exotic}}

In this section, we highlight topics of fundamental physics searches
that were discussed in recent years and whose scenarios could be
studied or hopefully constrained with CTA.

There are several caveats if one wants to present such various and
complex topics ``in a nutshell''. First of all, the list of topics is
not exhaustive;  only a subset of topics is reported here. Second,
some of the studies presented 
in this section may not be up-to-date by the time this article is
published: theories in this area evolve and are updated exceedingly
rapidly. In addition, most of these studies were formulated only
within the context of the current generation of IACTs, and not for
CTA. Whenever possible, considerations about the prospects for CTA
will be addressed. Third, the discussion will mostly be of only
a qualitative nature. The goal in this contribution is to provide an
introductory discussion of the area, with the aim of encouraging
others to explore in more detail these and other interesting new
physics possibilities. Let us add that pursuing exotic physics with
IACTs (and hence CTA) should be done because a) it is possible  
(this may seem a naive argument, but, given the {\em terra incognita}
offered by a new observatory such as CTA, it is a strong one; b) VHE gamma 
rays have been identified as likely drivers of truly fundamental
discovery. VHE gamma rays are a tool to explore new physics and new
astrophysical scenarios, the nature of which may contain yet
unknown, and unexpected, features. The potential for revolutionary
discovery is enormous. 

\subsection{High energy tau-neutrino searches\label{sec:tau}}

Although optimized to detect electromagnetic air showers produced by
cosmic gamma-rays, IACTs are also sensitive to hadronic
showers. Inspired by calculations made by the AUGER collaboration and
D.~Fargion~\cite[see,
  e.g.,][]{1999ICRC....2..396F,2002ApJ...570..909F,2001AIPC..566..157L,2002PhRvL..88p1102F,Abraham:2008,Bertou:2001vm},
the possible response of IACTs to showers initiated by very high
energy $\tau$-particles originating from a $\nu_\tau$ collision with
the sea or underneath rock is described.

It is well known that neutrinos of energies above the TeV energy range
can form part of the cosmic rays hitting the Earth. The origin of such
neutrinos could be from point-like sources like galactic
microquasars~\cite{Torres:2004tm, Bednarek:2005gf} or extragalactic
blazars~\cite{Mannheim:1998wp, Muecke:2002bi} or gamma-ray
bursts~\cite{2004APh....20..429G}. There are also diffuse fluxes of
high energy neutrinos predicted to come from unresolved sources,
including interactions of EHE cosmic rays during their
propagation~\cite{Berezinksy}. Finally, one could think of a more 
exotic origin of high energy neutrinos like those coming from DM
particle annihilation, topological defects or cosmic
strings~\cite{Witten:1985,Hill:1987,Berezinsky:2011cp}.
Neutrinos are produced in astrophysical sources or during the
transport, mainly after pion and subsequent muon decays:
\begin{eqnarray}
\pi^+\rightarrow\mu^+ \;\nu_\mu &\qquad & \mu^+\rightarrow\bar{\nu_\mu}\;e^+\;\nu_e\\
\pi^-\rightarrow\mu^- \;\bar{\nu_\mu}  &\qquad& \mu^-\rightarrow\nu_\mu\;e^-\;\bar{\nu_e}
\end{eqnarray}
such that the typical neutrino family mixing at the source is
$(\nu_e,\nu_\mu,\nu_\tau) = (1:2:0)$. Tau-neutrinos are found either
at the source, if charmed mesons are formed instead of pions, or are
created during the propagation, after flavor mixing, such that at
Earth, the neutrino family mixing could be 
$(\nu_e,\nu_\mu,\nu_\tau)_{\mbox{\small{Earth}}} =
(1:1:1)$~\cite{Learned:1995}.

The $\nu_\tau$ channel has several advantages with respect to the
electron or muon channel. First, the majority of the possible $\tau$
decay modes lead to an (observable) air shower or a combination of
showers. Only 17.4\% of the decays lead to a muon and neutrinos,
considered to be unobservable for the effective areas of interest
here. Moreover, the boosted $\tau$ lifetime ranges from some 50~m at
1~PeV to several tens of kilometers at EeV energies, almost unaffected
by energy losses in matter and thus surpassing the muon range by a
factor of $\sim$~20. Finally, the originating $\tau$ decays, instead
of being absorbed by matter, and thus gives origin to another
$\nu_\tau$ of lower energy which in turn can produce a $\tau$. At the
highest energies, the Earth becomes completely opaque to all types of
neutrinos giving rise to a pile-up of $\nu_\tau-\tau$.

\begin{figure}[h!t]
\centering
\includegraphics[width=0.95\linewidth]{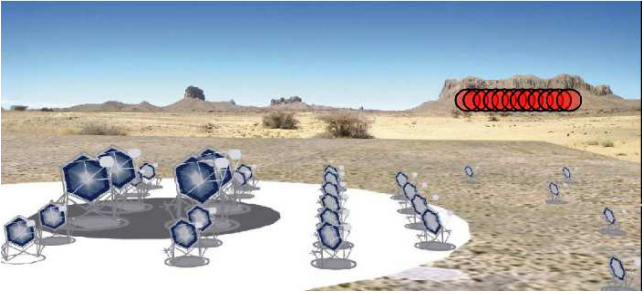}
\caption{\label{fig:tau} Sketch of $\nu_\tau$ searches with CTA. CTA
  telescopes can be pointed horizontally to a far mountain (if
  present) and observed the emerging $\tau-$induced atmospheric shower.}
\end{figure}

To be able to observe atmospheric showers from $\nu_\tau$, the
telescopes should be pointed at the direction where the $\tau$ escapes
from the Earth crust after having crossed an optimized distance inside
the Earth. Of course, this distance is strongly dependent on the
telescope location, and no general conclusions can be drawn before the
CTA site will be defined. In the past, two directions were
proposed\footnote{The study was made for the case of
  MAGIC~\cite{2008ICRC....3.1273G}}: the observation slightly below
the horizon downhill, e.g. if the telescope is located at a mountain, and the observation through a
possible mountain chain in the vicinity (see
Fig.~\ref{fig:tau}). Both observations are at 
extremely low elevation angles, i.e. the telescopes pointing
horizontally or even below the horizon, a fact that guarantees that
the hadronic background diminishes until almost vanishing at the
horizon. Only a small light contamination from continuous scattered
star light and from scattered Cherenkov light by air showers will then
be observed.

In Ref.~\cite{2008ICRC....3.1273G}, the effective area for $\nu_\tau$
observation with the MAGIC telescope was calculated analytically. 
The results were the
following: the maximum sensitivity would be in the range 100~TeV--1
EeV. For the observation downward towards the Sea, the sensitivity for
diffuse neutrinos is very low because of the limited FOV
(3~events/year/sr) and CTA cannot be competitive with other experiments
like Icecube~\cite{Abbasi:2012cu}, Baikal~\cite{Balkanov:1999},
Auger~\cite{Abraham:2008}, Antares~\cite{AdrianMartinez:2011uh} or KM3NeT. On the other hand, if flaring
or disrupting point sources are observed, like is the case for GRBs,
one can even expect an observable number of events from one GRB at
reasonable distances, if the event occurs just inside a small
observability band of about 1 degree width in zenith and an azimuth angle which allows to point the telescopes downhill.

For CTA, the situation could be different: taking an extension of the
FOV of several times that of MAGIC in extended observation mode,
the higher effective area and lower energy threshold, meaning higher
fluxes, one na\"ive rescaling of the MAGIC calculations leads to
relatively optimistic results, depending very much on the local
geography.  For point-like sources, the situation would not change so
much w.r.t. the MAGIC case, unless the CTA telescopes are located
close to a shielding mountain chain. The required observation times are still large, but one may argue that these observations can be
performed each time when high clouds preclude the observation of gamma-ray sources.

\subsection{Ultrarelativistic Magnetic Monopoles\label{sec:monopoles}}

The existence of magnetic monopoles is predicted by a wide class of
extensions of the standard model of particle
physics~\cite{Groom:1986}. Considerable experimental effort has
been undertaken during the last eight decades to detect magnetic
monopoles. No confirmed success in detection has been reported at the
present time. Current flux limits on cosmogenic magnetic monopoles reach values
of $\mathcal{O}(10^{-15}\textrm{
  cm}^{-2}\textrm{s}^{-1}\textrm{sr}^{-1})$ to
$\mathcal{O}(10^{-17}\textrm{
  cm}^{-2}\textrm{s}^{-1}\textrm{sr}^{-1})$ depending on the monopole
velocity. As outlined below, the CTA observatory is sensitive to a
magnetic monopole flux.

\begin{figure}[h!t]
  \centering
  \includegraphics[width=0.95\linewidth]{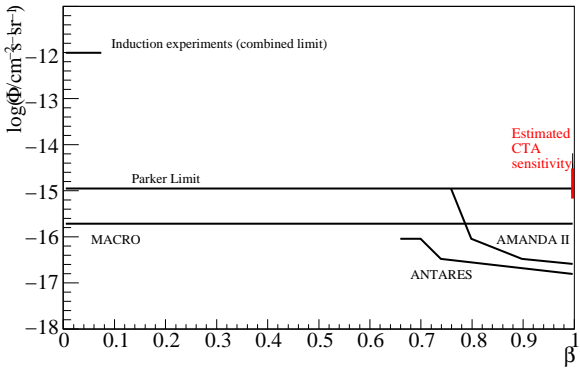}
  \caption{\label{fig:monopole}Upper limits on the magnetic monopole
    flux as function of the monopole velocity inferred from different
    experiments~\cite{Groom:1986, Achterberg:2010} together with
    the estimated sensitivity of CTA.  The sensitivity range shown for
    CTA indicates the dependence of the sensitivity on CTA design
    parameters and analysis details.}
\end{figure}

According to \citet{Tompkins:1965} magnetic monopoles moving in
air faster than the speed of light in air are emitting $\approx 4700$
times more Cherenkov photons than an electric charge under the same
circumstances. Being fast enough (Lorentz factor $\gamma>10^3$) and
heavy enough (mass $Mc^2>1\textrm{ TeV}$) magnetic monopoles that
possibly propagate through the earth atmosphere are neither
significantly deflected by the Earth's magnetic field nor loose a
significant amount of energy through
ionization~\cite{Spengler:2011}. Assuming the last two constraints to
be fulfilled a magnetic monopole moving through the Earth's atmosphere
propagates on a straight line, thereby emitting a large amount of
Cherenkov photons. This process of a uniform emission of intensive
Cherenkov light differs from the Cherenkov light emitted by secondary
particles in a shower initiated from a high energy cosmic or
gamma-ray. As shown by \citet{Spengler:2011}, the number of triggered
pixels in a telescope array is typically smaller and the intensity of
the triggered pixels is typically higher for magnetic monopoles
compared to events originating from cosmic or gamma-rays.  Cuts in a
parameter space spanned by the number of triggered pixels in the CTA
array and the number of pixel with high intensity allow for an
excellent discrimination between magnetic monopole events and
background from cosmic or gamma-rays. The effective detection area of
H.E.S.S.~\cite{Aharonian:2006crab} for magnetic monopoles has been
studied in detail~\cite{Spengler:2011}. Extrapolating the results of
this study for CTA with its one order of magnitude increased design
collection area, leads to a typical CTA magnetic monopole effective
area of $4500\textrm{ m}^2\textrm{sr}$. In Fig.~\ref{fig:monopole}, we
show that assuming around $3000$ hours of CTA data from
different observations accumulated in about $4$ years of array
operation, the sensitivity of CTA to magnetic monopoles with
velocities close to the speed of light can reach the Parker limit
\cite{Groom:1986} of $\mathcal{O}(10^{-15}\textrm{
  cm}^{-2}\textrm{s}^{-1}\textrm{sr}^{-1})$. Despite being still two
orders of magnitude worse than current monopole flux limits from
neutrino experiments \cite{Achterberg:2010} this sensitivity will
allow a technically independent and new test for the existence of
magnetic monopoles.

\subsection{Gravitational waves\label{sec:gw}}

The period of operation of CTA should hopefully see the detection of
the first gravitational wave (GW) by ground-based interferometers, now
in the ``advanced sensitivity'' design phase. The 3\,km--scale Michelson
interferometers Enhanced LIGO and Advanced Virgo~\cite{Accadia2011}
are increasing their sensitivity and extending the horizon distance of
detectable sources up to hundreds of Mpc, depending on the
frequency.
Two additional
smaller interferometric detectors are part of the network of GW
observatories, the Japanese TAMA (300\,m arms) and the German-British
600\,m interferometer~GEO600.
The LIGO and Virgo observatories should start full operation in the
advanced version in 2014/2015 and may operate together to fully
reconstruct the arrival direction of a signal. They may localize
strong GW bursts with an angular uncertainty down to one degree, while
weaker signal have larger uncertainties, up to tens of
degrees~\cite{Sylvestre:2003}.
At the stage of
conceptual design, the Einstein Telescope aims at increasing the arm
length to 10\,km, with three arms in a triangular pattern,
implementing consolidated technology. It is foreseen to be located
underground to reduce the seismic motion thus allowing a better
sensitivity up to a factor 10~\cite{Sathyaprakash:2011}.

The most promising astrophysical mechanisms able to produce observable
GWs are in--spiral and coalescence of binary compact objects (neutron
stars and black holes), occurring for example during the merger of
compact binary systems, as progenitors of supernova or neutron-star
collapse, and associated with pulsar glitches. Signals
from these systems may last from milliseconds to a few tens of
seconds, but their expected rate and strength are uncertain~\cite[for
  a review, see, e.g.,][]{Buonanno2007}. Moreover, unexpected or
unknown classes of sources and transient phenomena may be responsible
for GW emission and may actually provide the first detection.
Therefore combined GW and electromagnetic observations would be
critical in establishing the nature of the first GW detection. While
electromagnetic counterparts cannot be guaranteed for all GW
transients, they may be expected for some of them~\cite{Bloom2009,
  Chassande:2011} from radio waves to gamma-rays, such as in gamma ray
bursts \cite{Stamatikos2009}, ultra high-luminous X-ray transients and 
soft gamma repeater \cite{Abbott2008}. Electromagnetic identification
of a GW would confirm the GW detection and improve the reconstruction
and modeling of the physical mechanism producing the event.  Moreover,
significant flaring episodes identified in the electromagnetic band
could serve as an external trigger for GW signal identification, and
could even be used to reconstruct independently the source position
and time, thus allowing the signal-to-noise ratio required for a
confident detection to be lowered. The feasibility of this approach
has been corroborated through dedicated simulations by the LIGO and
Virgo collaborations~\cite{Abadie:2011}.

CTA has the capability to pursue such a program of immediate follow-up
of target of opportunity alerts from GW observatories, and to interact
with GW collaborations to pursue offline analysis on promising
candidates. The capability of CTA to observe in pointed mode with
small FOV or in extended mode covering many square degrees
of sky, is unique to follow strong and weak GW alerts. The observation
mode should resemble the GRB procedure, which allow a fast
repositioning on the order of tens of seconds.

In the era in which ground-based gravitational wave detectors are
approaching their advanced configuration, the simultaneous operation
of facilities like CTA and Virgo/LIGO may open, in the forthcoming
years, a unique opportunity for this kind of multi-messenger search.

%
%
\section{Summary and Conclusion}
\label{sec:conclusion}

In this study we have investigated the prospects for detection and
characterization of several flavors of physics beyond the standard
model with CTA.\\

\subsection*{Particle Dark Matter searches}
We have investigated dark matter (DM) searches with CTA for different observational
strategies: from dwarf satellite galaxies (dSphs) in
Section~\ref{sec:dwarf}, from clusters of galaxies in
Section~\ref{sec:gc} and from the vicinity of the Galactic Centre
in Section~\ref{sec:halo}. In Section~\ref{sec:spatial}, we discussed
spatial signatures of DM in the diffuse extragalactic gamma-ray
background.

Concerning searches in dSphs of the Milky Way, we have investigated
the prospects for detection of well-known ``classical'' dSph like Ursa
Minor and Sculptor, and one of the most promising ``ultra-faint''
dSph, Segue~1 (Table~\ref{tab:jbar}). We have first shown that the
predictions for core or cusp DM density profiles are quite similar for
the baseline CTA angular resolution (Fig.~\ref{fig:profiles}). We have
then simulated a 100~h observation for several CTA arrays, and found
that for Segue~1, we can exclude velocity-averaged cross-sections
$\left(\sigma_{\rm ann} v\right)$ above
$10^{-23}-10^{-24}\,\mbox{cm}^3\mbox{s}^{-1}$ depending on different
annihilation channels (Fig.~\ref{fig:sigmav_ul}). We also presented
the same results in terms of the minimum astrophysical factor for dSphs
to be detected (Fig.~\ref{jfactors}), showing that astrophysical
factors of at least $10^{21}$~GeV$^{2}$~cm$^{-5}$ are needed. We finally
showed the minimum intrinsic boost factor to achieve detection
(Fig.~\ref{fig:Boost}), which for Segue~1 is about 25 for a hard
annihilation spectrum. The best candidate arrays for dSph study are
array $B$ and $E$. Nevertheless, the robustness of our results is hindered
by the yet not precise determination of the astrophysical factor in
some cases.  Forthcoming detailed astronomical measurements will
provide clues for deep exposure observations on the most promising
dSphs, with, e.g., the planned SkyMapper Southern Sky
Survey~\cite{Keller:2007}, which will very likely provide the
community with a new dSph population, complementing the Northern
hemisphere population discovered by the SDSS. Also, the uncertainties
on dark matter density will be 
significantly reduced by new measurements of individual stellar
velocities available after the launch of the GAIA
mission\footnote{www.rssd.esa.int/Gaia}.
Stacking-methods of Fermi-LAT dSphs data were proven valid to make
constraints more stringent~\cite{Abdo:2010b,GeringerSameth:2011iw,
  Cotta:2011pm}. The application of these methods for CTA is currently
under study.

The search for DM signatures in galaxy clusters, investigated in
Section~\ref{sec:gc} was performed for two representative clusters,
Perseus and Fornax. The former one is thought to have the highest
CR-induced photon yield, and the latter is thought to have the
strongest DM-induced signatures. Compared to dSphs, the gamma-ray
signatures of galaxy clusters have several contributions: in the first
place, the DM signal is expected from an extended region that can be
larger than a few degrees, and secondly, gamma-rays induced by
interactions of accelerated cosmic rays with the ambient fields and/or by
individual cluster galaxies are an irreducible background to the DM
signal, as recently shown in Refs.~\cite{Aleksic:2010xk,
  Aleksic:Perseus}. We have simulated the prospects of detection in
100~h of observation by using MC simulations of extended sources.
Regarding DM signatures, we have used the model
of~\citet{Pinzke:2011ek} for the Fornax cluster, and showed that in
100~h we could put contraints on the order of $\left(\sigma_{\rm ann}
v\right)<10^{-25}\,\mbox{cm}^3\mbox{s}^{-1}$
(Fig.~\ref{fig:Fornax_DM}), which are competitive with respect to
those obtained with dSphs. The results are promising: if the intrinsic
boost factor from subhalos is larger than that predicted by the
model we used, or mechanisms of Sommerfeld enhancement are at work,
there is also the possibility to have a detection in $100-200$~h with
array $B$ or $E$.  We have also considered the prospects of detection of
CR-induced signal in hadronic acceleration scenarios in
Fig.~\ref{fig:Perseus_CR}. We have seen that the CR-induced emission
from the Perseus cluster could be detected in about 100~h.  Finally, we
discussed the more realistic case when DM-- and CR-induced gamma-rays
are treated together. We discuss that the difference in both the
spatial and spectral features of the two emissions can be used as a
method for discrimination, while more quantitative results need
dedicated MC which were not available when writing this contribution.
We underline that the extension of the expected DM emitting region in
galaxy clusters represents a problem for current Cherenkov Telescopes
since their FOV is limited to $3-5$ degrees and their
sensitivity rapidly decreases moving away from the centre of the
camera. CTA will overcome this limitation, having a FOV of
up to 10~deg and an almost flat sensitivity up to several degrees from
the centre of the camera. For galaxy cluster searches, CTA will hence
mark the difference compared to the current generation of IACTs.

More promising are DM searches of annihilation signatures in
the Galactic halo, where the DM density is expected to be known with
much higher 
precision than in the Galactic Centre itself or in (ultra-faint) dSphs
or galaxy clusters. This was studied in Section~\ref{sec:halo}. By
adopting dedicated observational strategies of the region close to the
Galactic Centre, as shown in Fig.~\ref{fig:halo_rings}, it was shown
that CTA has the potential to reach the thermal annihilation cross-section expected from WIMP DM of
$10^{-26}\,\mbox{cm}^3\mbox{s}^{-1}$ and lower
(Fig.~\ref{fig:halo_methods_spectra}) in 100~h observation of the
vicinities the Galactic Centre using the ``Ring'' method. Models
with a large photon yield from DM annihilation will be constrained for
even smaller cross-sections. It is also expected that the limits
presented here can be improved by factor of a few when the stereoscopic
analysis of CTA events has been understood so well that a further
suppression of the background becomes feasible. This would be the first time that
ground-based Cherenkov telescopes could reach this sensitivity level.

Besides observations of individual dedicated objects, the capabilities
of CTA for searching DM signals in the diffuse background of gamma-ray
radiation were discussed in Section~\ref{sec:spatial}. We discussed
the reconstruction performance for different anisotropy power spectra
and residual background level. Considering a current model for the
anisotropy power spectra, we showed that CTA may be able to distinguish
a DM-induced diffuse gamma-ray component from the astrophysical
background.\\ 


\begin{figure}[h!t]
  \centering
  \includegraphics[width=0.95\linewidth]{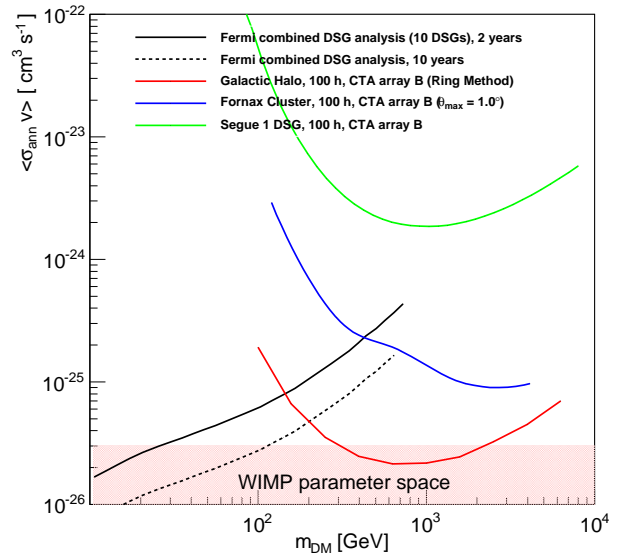}
  \caption{\label{fig:Fermi.vs.CTA} Comparison of exclusion curves of
    Fermi-LAT in 24~months \cite{Ackermann:2011} and expected for
    10~years (rescaled with the square root of time). The exclusion
    curves for the various targets studied in this contribution are
    also reported for the $b\bar{b}$ annihilation channel: for the
    dwarf satellite galaxy Segue~1 (green curve, see
    Sec.~\ref{sec:dwarf}), for the Fornax galaxy cluster in case only
    DM-induced gamma-rays are considered (blue line, see
    Sec.~\ref{sec:gc}) and for the ring-method of observation of the
    Galactic Centre vicinities (red line, see Sec.~\ref{sec:halo}). }
\end{figure}

In Fig.~\ref{fig:Fermi.vs.CTA}, we summarize the constraints that we
expect with CTA for a WIMP annihilating purely into $b\bar{b}$ in
100~h observation, with the different targets discussed above.  As
already anticipated, the best results are expected for the observation
of the vicinity of the Galactic Centre, where we expect to reach the
thermal annihilation cross-section for
WIMP DM of $10^{-26}\,\mbox{cm}^3\mbox{s}^{-1}$.  Unlike present
IACTs, whose sensitivity supersedes that of the Fermi-LAT at masses
around a TeV, CTA will
constitute the most sensitive instrument above masses of about
100~GeV. It should be noted that these estimates are conservative: the
most important improvement can be expected from the possible
redefinition and final optimization of the array layout\footnote{For
  example, it is currently under disussion, the possibility to add 36
  medium-size telescopes of Schwarzchild-Couder design to the arrays
  considered here. Preliminary simulations predict that improvement in
  the overall sensitivity by at least a factor of 2 compared to that
  studied here may be expected.}  In addition, the presented
sensitivities were calculated using generic analysis which was not
optimized specifically for the DM searches and thus our results could
be considered conservative in this sense.

Obviously, a firm \emph{identification} of DM requires a very good spectral
discrimination with respect to any possible astrophysical
background. Spectral shapes and (even more so) absolute normalization
of these backgrounds are often poorly determined and the DM signal
most likely is small in comparison. As the extent to which these
factors affect detection claims is highly model and target dependent,
we refer to future detailed more focused assessments.\\

It has been shown that the detection of gamma-rays provides complementary
information to other experimental probes of particle DM, especially
that of direct detection, because CTA could be able to access a
fraction of the parameter space not accessible
otherwise~\cite{Bergstrom:2010gh, Bergstrom:2011}. With respect to
particle searches at the LHC, the comparison is not straightforward,
as LHC results are usually strongly related to specific models, and
general conclusions are somewhat model dependent, as shown by recent
publications from the ATLAS and CMS
collaborations~\cite{Khachatryan:2011tk, Aad:2011ks,
  Aad:2011hh}. Generically, the discovery of a candidate for particle
DM will be limited by the available centre-of-mass energy. Other
scenarios exist, in the context of specific super-symmetric models for DM,
that exhibit parts of the model space not accessible by
the LHC~\cite{Profumo:2011zj}. In any case, LHC discovery of dark matter,
would prompt the need for proof that the particle is actually consistent with the
astrophysical DM, and close collaboration with LHC physicists is
currently under organization to facilitate the optimal use of
accelerator results within CTA. A concrete scenario has been analyzed
by~\citet{Bertone:2011pq} in the case of a SUSY model in the so-called
co-annihilation region. Simulated LHC data were used to derive
constraints on the particle physics nature of the DM, with the result
that the LHC alone is not able to reconstruct the neutralino
composition. The situation improves if the information from a
detection of gamma-rays after the observation of the Draco dSph by CTA
is added to the game: in this case the internal degeneracies of the
SUSY parameter space are broken and including CTA allows us to fully
interpret the particle detected at the LHC as the cosmological DM.  
In
the other case where the LHC will not detect any physics beyond the
Standard Model, predictions were
made in the context of the CMSSM~\cite{Bertone:2011kb} indicating that
the mass of the neutralino will be bound to be larger than
approximately $250$~GeV ($400$~GeV) if any new physics will be
detected by the LHC for an energy of the centre-of-mass
$\sqrt{s}=14$~TeV and a luminosity of $1$~fb$^{-1}$
(100~fb$^{-1}$). In this scenario, CTA could be the only
instrument to be able to detect and identify a WIMP candidate with
masses beyond some hundreds GeV.

\subsection*{Axion-like particle searches}
In Section~\ref{sec:axions}, the prospect of searches for axion-like
particle (ALP) signatures with CTA were studied. We saw that the
theoretical photon/ALP mixing has important implications for
astronomical observations, in such that the mixing could distort the
spectra of gamma-ray sources, such as Active Galactic Nuclei (AGN) (or
galactic sources), in the TeV range. This distortion adds to that
caused by the absorption of the gamma-ray photons with UV and IR
photons of the Extragalactic Background Light (see
Fig.\ref{fig:sketch}).  The photon flux recently measured by some
experiments, in particular at TeV energies, already exceeds that
predicted by conventional models which attempt to explain spectra in
terms of  observed source spectra and/ or EBL density~\cite{aleksic11,
  aliu09, Nesphor1998, Stepanyan2002, Krennrich2008}, one
should not expect a photon flux as high as recently measured by some
experiments, in particular at TeV energies. The hard spectrum deduced
for some AGN is difficult to explain with conventional physics as
well. While it is still possible to solve these puzzles without exotic
physics, photon/ALP conversions may naturally alleviate both problems.
In order to quantitatively study the effect of photon-axion conversion
over cosmological distances, the total photon flux from a simulated
flare of a far-distant source was considered. The source was simulated
based on the flat spectrum radio quasar 4C +21.35 (PKS 1222+21, $z =
0.432$, based on the observation performed by
MAGIC~\cite{pks1222_MAGIC}), assuming an intrinsic unbroken power-law
spectrum, and trying to understand the observability under different
absolute flux normalization and flare duration
(Figs.~\ref{fig:Ecrit200} and \ref{fig:Ecrit1000}).  The range of
characteristic scale energy (critical energy, $E_{crit}$) described in
Eq.~(\ref{eq:ecrit}), and thus the ALP mass that can be probed with CTA
for the different ALP scenarios, is unknown and may be tested with
CTA. In general, the distortion of the spectra due to ALP depends on
the particular case, but as a general trend it will become larger as
we increase the observation time and/or the flux of the source. As an
example, we found that a 0.5 h duration flare like the one reported by
MAGIC would not be enough for CTA to detect a significant effect in
any of the tested ALP scenarios
(Fig.~\ref{fig:probabilities}). However, a flare of similar intensity,
but lasting 5 h would already be enough to see the boost due to ALPs
for those scenarios with E$_{crit} \leq 500$ GeV. For a hypothetical
flare with an intensity 5 times larger, lasting 5 hours, the
accessible range of E$_{crit}$ would extend up to 1.3 TeV
(Fig.~\ref{fig:probabilities}). Hopefully, not only PKS 1222+21 but
also many other similar objects will be followed-up by CTA in the near
future, making the field of ALP searches very promising.

We must emphasize that a boost in the flux is {\it only} possible in
the energy range where the EBL is already at work. Thus, even for the
most distant sources detected to date by IACTs, the energy range below
$\sim100$ GeV would not probably be of much help. On the other hand,
even when $E_{crit}$ lies within the energy range covered by IACTs,
the drop/jump might not be accessible to these instruments. This would
be the case, for instance, if $E_{crit}$ is at the highest energies,
from several to tens of TeV: the attenuation due to the EBL for a
distant source would be huge, and the resulting flux, even after
accounting for the ALP boost, too low to be detected by current IACTs
or by CTA under any of the possible array configurations. Taking all
these considerations into account, the most suitable energy for ALP
searches with CTA seems to be an intermediate one in which the EBL is
already present but still introduces only a moderate absorption,
i.e. from a hundred GeV to a few TeV. As a result, we do not expect to
obtain largely different results with candidate array configurations
other than the one we used (array $E$), since they all perform very
similarly in the intermediate energy range.

Finally, although very challenging given the uncertainty in the value
of the IGMF, we should stress that the lack of detection of suspicious
features in the spectra of distant gamma-ray sources might translate
into useful constraints of the ALP parameter space (coupling constant
and ALP mass). A more detailed study is definitely needed in order to
find out what should be the best strategy to achieve the strongest
constraints. This study will be done elsewhere.

\subsection*{Lorentz Invariance Violation}
In scenarios where Lorentz invariance is violated by quantum
gravitational effects, the space-time fabric may be distorted so that 
the vacuum shows a non-unitary refractive index and thus the light
speed would be wavelength dependent. Observation of gamma-ray flares
from far distant objects like active galactic nuclei or gamma-ray
bursts, may allow to detect the time--delay between photons of
different energies not caused by intrinsic source mechanisms.  In
Section~\ref{sec:liv}, we discussed the
sensitivity of the different CTA array configurations on detecting time--delays
induced by Lorentz Invariance Violations (LIV). While
limits on LIV from the current generation of IACTs are weaker than
those estimated from Fermi-LAT measurement in the so-called linear
case, CTA is likely to invert this scenario.
 
Using for the Maximum Likelihood Estimation
method of \citet{martinez09}, 500 Gaussian-shaped pulsed light-curves
with time--delay from $-60$~s TeV$^{-1}$ to $60$~s TeV$^{-1}$ were
simulated and reconstructed with the different arrays. CTA will have
improved statistics of photons and larger spectral lever-arm due to
the enlarged energy range with respect to the current generation of
IACTs. This will allow to better differentiate between the two
Gaussian peaks as shown in Fig.~\ref{fig:jb1} for the different
arrays. This ability to differentiate peaks is also discussed in
Fig.~\ref{fig:jb2} for different width of the peaks. The best array
configurations were discussed. As a result of these studies, a gain of
about a factor 50 in the LIV scale for the ``quadratic'' model (see
Eq.~(\ref{eq:liv})) is expected, compared to current generation of
telescopes, while the limits on the linear term will largely exceed
the Planck energy scale.

We then used extrapolation to high-energies of real
AGN spectra observed by the current generation of IACTs for three
representative scenarios: a very bright AGN (Mrk~421), a fast-variable
one (PKS~2155-304) and a high-redshift one (3C~279). High-energy
photons above 10~TeV will guarantee the best sensitivity to observe
LIV signatures, and CTA with its improved sensitivity at those high
energies, will allow to collect sufficient photons, whereas photon
statistics will always be the final limiting factor on tests for
time--delay.

Finally, pairs of realistic AGN lightcurves with and
without time--delay were simulated, and folded with CTA
performance. For each array, we calculated the fraction of photons in
which the time--delay was successfully measured according to a quality
factor $q$ (Eq.~\ref{eq:qfactor}). In the most stringent case $(q<1)$,
we report the photon fraction recovery of each individual array in
Fig.~\ref{fig:globalAGNHist}. As a main result, more than 10\% of the
time--delays can be recovered with several possible CTA arrays. If we
relax the quality factor, and thus the precision on the reconstructed
time--delay, essentially all the time--delays are recovered. We showed
that arrays $C$, $D$, $H$, $I$ and $NB$, respectively for the Southern
and Northern hemisphere, have the best chance for detection.

Based on these genuinely different time--delay reconstruction methods we
ensure that our final results, with respect to CTA-array ranking, are
free from any possible systematic effects related to a given analysis
method e.g.\ idealized source redshift-distribution, idealized source
light-curves, and idealized source time-scales. In all analyses
methods the sub-arrays $C$, $H$, $I$ and $NB$ seem to be sufficiently good to
perform detection of LIV effects by measuring differences in the
arrival times of VHE photons. That means that these arrays are in
general sufficiently good to perform temporal
studies of light-curve signals and even detection of time--delays in
AGN induced intrinsically in the source. The latter is an interesting
degeneracy, connected with the actual origin of the time--delays, that
CTA will definitely be able to break through population studies not
based on exceptional flaring states but on a routine basis.

\subsection*{Other Searches}
Finally, in Section~\ref{sec:exotic}, we have qualitatively discussed
the physics case of a selection other exotic physic searches which are in principle
possible with CTA: the observation of atmospheric showers from
$\tau$-particles emerging from the Earth crust, the observation of
atmospheric showers from magnetic 
monopoles, and the possible follow-up of gravitational waves
events. Despite the prospects being sometimes pessimistic, those subjects were shown to
underline again the possibility of using an astronomical observatory
such as CTA for fundamental physics searches.\\

As a final closing remark, we believe that CTA could offer one of the
most powerful tools in the study of some of the most pressing
questions in modern physics. In the next few years it may lead to a
range of new observables, new methods and new theories. In preparation
for these developments, it is essential that work such as that
performed here is continued.

\subsection*{Acknowledgment}
\small{We gratefully acknowledge support from the following agencies and organisations:
Ministerio de Ciencia, Tecnolog\'ia e Innovaci\'on Productiva (MinCyT),
Comisi\'on Nacional de Energ\'ia At\'omica (CNEA) and Consejo Nacional  de
Investigaciones Cient\'ificas y T\'ecnicas (CONICET) Argentina; State Committee
of Science of Armenia; Ministry for Research, CNRS-INSU and CNRS-IN2P3,
Irfu-CEA, ANR, France; Max Planck Society, BMBF, DESY, Helmholtz Association,
Germany; MIUR, Italy; Netherlands Research School for Astronomy (NOVA),
Netherlands Organization for Scientific Research (NWO); Ministry of Science and
Higher Education and the National Centre for Research and Development, Poland;
MICINN support through the National R+D+I, CDTI funding plans and the CPAN and
MultiDark Consolider-Ingenio 2010 programme, Spain; Swedish Research Council,
Royal Swedish Academy of Sciences financed, Sweden; Swiss National Science
Foundation (SNSF), Switzerland; Leverhulme Trust, Royal Society, Science and
Technologies Facilities Council, Durham University, UK; National Science
Foundation, Department of Energy, Argonne National Laboratory, University of
California, University of Chicago, Iowa State University, Institute for Nuclear
and Particle Astrophysics (INPAC-MRPI program), Washington University McDonnell 
Centre for the Space Sciences, USA.

We thank I.~Freire,
  S.~Funk, W.~Hofmann, A.~Murphy, A.~Pinzke, S.~Sarkar, D.~Torres and
  F.~Zandanel who provided 
  comments on the manuscript.  D.~Emmanoulopoulos acknowledges the
  Science and Technology Facilities Council (STFC) for support under
  grant ST/G003084/1. This work was partially supported by the Spanish
  Consolider-Ingenio CPAN (CPAN09-PD13) and Multidark
  (CSD2009-00064). A. Jacholkowska and J. Bolmont acknowledge the support of GdR PCHE in France.

\bibliographystyle{model1-num-names}
\bibliography{bib_total}

\end{document}